\documentclass[a4paper,11pt]{article}

\usepackage{booktabs}
\usepackage{footnote}
\usepackage{jcappub} 
\usepackage{lineno}
\usepackage{graphicx} 
\usepackage{hyperref}
\usepackage{bm}
\usepackage{bbm} 
\usepackage{algpseudocodex} 
\usepackage{algorithm} 
\usepackage{subcaption} 
\usepackage{dcolumn} 
\usepackage{enumitem}
\usepackage{multirow} 
\usepackage{ragged2e}
\usepackage{siunitx} 
\usepackage[draft,author=,nomargin,inline]{fixme}

\makesavenoteenv{table} 
\fxusetheme{color}
\FXRegisterAuthor{imv}{aimv}{Ivan}
\FXRegisterAuthor{cfs}{acfs}{Carlos}

\DeclareSIUnit{\solarmass}{\ensuremath{\mathit{M_{\odot}}}}
\DeclareSIUnit{\parsec}{pc}

\renewcommand{\vec}[1]{{\bf #1}}
\newcommand{\vecg}[1]{\bm{#1}}
\newcommand{\mat}[1]{#1}

\newcommand{\conj}[1]{#1^{\ast}}
\newcommand{\fourier}[1]{\tilde{#1}}

\newcommand{\dee}{\mathrm{d}} 

\newcommand{\order}{\mathcal{O}}
\newcommand{\normal}{\mathcal{N}}
\newcommand{\identity}{\mathbbm{1}}

\title{Efficient Massive Black Hole Binary parameter estimation for LISA using Sequential Neural Likelihood}

\author{Iv\'an Mart\'in~V\'ilchez}
\author{Carlos F. Sopuerta}

\affiliation{Institut de Ci\`encies de l'Espai (ICE,~CSIC), Campus UAB, Carrer de Can Magrans~s/n, Cerdanyola del Vall\`es~08193, Spain}
\affiliation{Institut d'Estudis Espacials de Catalunya (IEEC), Edifici RDIT, C/ Esteve~Terradas, 1, desp.~212, Castelldefels~08860, Spain}

\emailAdd{imartin@ice.csic.es}
\emailAdd{carlos.f.sopuerta@csic.es}

\date{\today}

\abstract{
The inspiral, merger, and ringdown of Massive Black Hole Binaries (MBHBs) is one the main sources of Gravitational Waves (GWs) for the future Laser Interferometer Space Antenna (LISA), an ESA-led mission in the implementation phase. It is expected that LISA will detect these systems throughout the entire observable universe.
Robust and efficient data analysis algorithms are necessary to detect and estimate physical parameters for these systems.
In this work, we explore the application of Sequential Neural Likelihood, a simulation-based inference algorithm, to detect and characterize MBHB GW signals in synthetic LISA data. We describe in detail the different elements  of the method, their performance and possible alternatives that can be used to enhance the performance. Instead of sampling from the conventional likelihood function, which requires a forward simulation for each evaluation, this method constructs a surrogate likelihood that is ultimately described by a neural network trained from a dataset of simulations of the MBHB signals and noise.
One important advantage of this method is that, given that the likelihood is independent of the priors, we can iteratively train models that target specific observations in a fraction of the time and computational cost that other traditional and machine learning-based strategies would require. 
Because of the iterative nature of the method, we are able to train models to obtain qualitatively similar posteriors with less than 2\% of the simulator calls that Markov Chain Monte Carlo methods would require.
We compare these posteriors with those obtained from Markov Chain Monte Carlo techniques and discuss the differences that appear, in particular in relation with the important role that  data compression has in the modular implementation of the method that we present. 
We also discuss different strategies to improve the performance of the algorithms.
}


\arxivnumber{2406.00565} 

\begin{document}

\maketitle

\flushbottom

\section{Introduction}
The first direct detection of gravitational waves (GWs) from a binary black hole merger by the Laser Interferometer Gravitational-Wave Observatory (LIGO) in 2015~\cite{LIGOScientific:2016aoc} has kick-started a new era in astronomy, where GWs constitute a new cosmic messenger that is providing invaluable information for astrophysics, cosmology, and even fundamental physics.
GW astronomy is advancing fast, as shown by the number of detections (more than 90 published so far) that have been achieved by the second generation of ground-based interferometric detectors~\cite{LIGOScientific:2014pky,VIRGO:2014yos,Aso:2013eba} and
presented by the LIGO-Virgo collaboration first~\cite{LIGOScientific:2016aoc,LIGOScientific:2018mvr,LIGOScientific:2021usb} and, more recently, by the LIGO-Virgo-KAGRA (LVK) collaboration~\cite{KAGRA:2021vkt}, including the first multi-messenger observation involving GWs~\cite{LIGOScientific:2017vwq,LIGOScientific:2017ync}.
Additional events outside the LVK collaboration have been reported in~\cite{Nitz:2018imz, Nitz:2020oeq, Nitz:2021uxj, Nitz:2021zwj,Olsen:2022pin,Mehta:2023zlk}.
All these observations, in the high-frequency GW band, have already provided revolutionary discoveries with impact in astrophysics, cosmology and even fundamental physics.
In parallel, in the very-low-frequency GW band (around the nanohertz), pulsar timing array collaborations have recently reported evidence for a GW background~\cite{NANOGrav:2023gor,EPTA:2023fyk,Reardon:2023gzh,InternationalPulsarTimingArray:2023mzf}.
There are also important developments towards the measurement of the cosmic microwave background polarization in order to put constraints on primordial GWs (see, e.g.~\cite{BICEP:2021xfz}).

On the other hand, due to the presence of various sources of noise, the sensitivity of ground-based detectors is limited to the frequency range between $10\,$Hz and $10\,$kHz~\cite{Abbott:2016xvh}, mainly due to seismic and (Newtonian) gravity gradient noises in the low end, and to quantum noise in the high end.
Third generation ground-based detectors like Einstein Telescope~\cite{Punturo:2010zz} and Cosmic Explorer~\cite{Reitze:2019iox} can push the low limit further up to $1\,$Hz, while atom-interferometric detectors~\cite{Badurina:2019hst,Canuel:2019abg} have been proposed to push it up to the decihertz.
Below this, the best option we currently have is to go to space.
The Laser Interferometer Space Antenna (LISA)~\cite{LISA:2017pwj} is a space mission led by the European Space Agency (ESA), in collaboration with the US National Aeronautics and Space Administration (NASA), that will survey the observable universe using low-frequency GWs.

LISA has recently achieved a milestone after passing the Mission Adoption Review and entering the implementation phase~\cite{LISA:2024hlh}, with a launch expected around 2035.
Prior to this, ESA approved the science theme for a GW space observatory, \emph{The Gravitational Universe}~\cite{eLISA:2013xep}, and commissioned a demonstration mission, LISA Pathfinder, which exceeded expectations by showing a noise differential acceleration performance at the level required for LISA~\cite{Armano:2016bkm,Armano:2018kix} (see~\cite{LISAPathfinder:2024ucp} for a detailed analysis of the LISA Pathfinder data and results).
The LISA mission consists of a constellation of 3 spacecraft in heliocentric orbit, forming an almost equilateral triangular array with an arm length of $2.5\times 10^6$ km.
Each spacecraft hosts two free-falling test masses and the relative distance between different spacecraft test masses is monitored using (time-delay) laser interferometry \cite{Tinto:2020fcc,Tinto:2022zmf}.

In this way, LISA will be sensitive to GW sources in the millihertz band, where we expect to detect GW signals from (see, e.g.~\cite{LISA:2017pwj,LISA:2024hlh}): (i) The coalescence of massive black hole (MBH) binaries (MBHBs), with total mass in the range $10^5-10^8\,M_\odot$; (ii) the capture and inspiral of a stellar-mass compact object into a MBH, known as an \emph{Extreme-Mass-Ratio Inspirals} (EMRIs), and similar systems known as \emph{Intermediate-Mass-Ratio Inspirals} (IMRIs) involving intermediate-mass black holes (IMBHs)\footnote{By definition, IMBHs have masses in the range $10^2-10^5\,M_\odot$. Then, we can distinguish between \emph{light} IMRIs, consisting in the inspiral of an stellar-origin black hole (SOBH) into an IMBH, and \emph{heavy} IMRIs, consisting of an IMBH inspiralling into a MBH.}; (iii) millions of ultra-compact galactic binaries (GBs), whose population is dominated by white dwarf binaries; (iv) Stellar-origin black hole (SOBH) binaries (like GW150914~\cite{LIGOScientific:2016aoc}) in the stage when their periods are in the $1\,$hour range; (v) stochastic GW backgrounds of cosmological origin, associated to high-energy processes at the TeV energy scale, and; (vi) given that LISA is expected to open the low-frequency band, we may find unforeseen GW sources for which we should be prepared.

By the detecting these GW sources, LISA will be able to carry out a wide science program~\cite{eLISA:2013xep,LISA:2017pwj,LISA:2024hlh} with revolutionary implications for astrophysics~\cite{LISA:2022yao}, cosmology~\cite{LISACosmologyWorkingGroup:2022jok} and fundamental physics~\cite{Barausse:2020rsu,LISA:2022kgy}.
The realization of the science program requires the development, prior to the beginning of the science operations, of the necessary computational infrastructure and methodology for the processing and scientific analysis of the LISA data.
In this sense, it is important to mention that LISA will be a signal-dominated detector in the sense that, at all times, we will find overlapping GW signals.
This requires data analysis algorithms that can simultaneously find and extract the parameters of all the sources, what is known as a \emph{global fit} pipeline (see~\cite{Littenberg:2023xpl,Strub:2024kbe,Katz:2024oqg} for recent proposals towards LISA global fit pipelines).

Many of the developments in the design of algorithms for LISA data analysis has taken place around a series of exercises with synthetic data, first under the name of Mock LISA Data Challenges~\cite{Arnaud:2007vr,Arnaud:2007jy,MockLISADataChallengeTaskForce:2007iof,Babak:2008aa,MockLISADataChallengeTaskForce:2009wir} and more recently, under the organization of the the LISA Data Challenge (LDC) working group of the LISA Consortium, established in 2018~\cite{LISADataChallenge2018} (see also~\cite{Baghi:2022ucj}).
In this way, the LDCs serve as a common benchmark to develop, test and compare different data analysis algorithms.

The current state of the art in LISA data analysis consists of using Bayesian sampling methods to draw samples from the posterior distribution $p(\vecg{\theta}\mid\vec{x})$, which is in turn proportional to a likelihood function $p(\vec{x}\mid\vecg{\theta})$ (here $\vecg{\theta}$ denotes the GW model parameters and $\vec{x}$ the detector data).
Among the most popular ones we have Markov Chain Monte Carlo (MCMC) sampling methods~\cite{Christensen:1998gf,Christensen:2001cr,Cornish:2005qw} (see~\cite{Cornish:2020vtw,Katz:2020hku,Marsat:2020rtl,Littenberg:2023xpl,Katz:2021uax} for recent applications to LISA) and nested sampling methods~\cite{Feroz:2009de,Gair:2010zz} (for recent applications to LISA, see~\cite{Spadaro:2023muy,Weaving:2023fji,Hoy:2023ndx}\footnote{Refs.~\cite{Weaving:2023fji,Hoy:2023ndx} adapt \texttt{PyCBC}~\cite{PyCBC} and \texttt{bilby}~\cite{bilby_paper} respectively for their use in LISA data analysis. They perform nested sampling by using their interfaces to the \texttt{dynesty}~\cite{Speagle:2019ivv,dynesty} sampler.}), but there are other options (see, e.g.~\cite{Jan:2024zhr}).
These methods, while generally reliable, require the repeated evaluation of the likelihood, which in turn demands a large number of waveform simulations during the inference process.
Despite the fact that they can be optimized for the problem in question, this ultimately results in high computational costs.
Moreover, the likelihood function typically used in GW data analysis comes from the assumption that the noise is (wide-sense) stationary and (multi-variate) Gaussian, so that it is completely described by the power spectral density (PSD), for which it is also assumed that we have an accurate model.
However, this does not reflect reality, indicating that likelihood-based methods will need to either acknowledge the systematic errors introduced, or to try to take them into account by adding corrections to the likelihood -- further driving up the computational cost.

The main goal of this work is to propose, as an alternative approach to likelihood-based methods, the use of simulation-based inference (SBI) techniques \cite{Cranmer:2019eaq}.
SBI leverages the recent developments in Machine Learning to obtain the posterior distribution using only the information contained within simulated data.
All assumptions about the data -- aside from the prior -- are thus built into the simulator itself.
As such, there is no need to define a likelihood function, making SBI applicable to problems where it is not possible to define a likelihood function or where it is numerically expensive to compute it. Moreover,
SBI techniques have already seen remarkable success when applied to GW source parameter estimation in the case of ground-based detectors (see, e.g.~\cite{Gabbard:2019rde,Bhardwaj:2023xph}).
Most notably, DINGO~\cite{Dax:2021tsq, Dax:2021myb, Dax:2022pxd} is able to do \emph{amortized inference}: after an initial training, a full inference run on previously unseen data with state-of-the-art accuracy can be performed in seconds to minutes, instead of hours to months.

In this paper, we explore the application of Sequential Neural Likelihood (SNL) estimation~\cite{Papamakarios:2018zoy} to LISA data analysis.
In brief, SNL replaces the manually-defined likelihood function, $p(\vec{x}\mid\vecg{\theta})$, with an approximate one, $p_{\vecg{\varphi}}(\vec{x}\mid\vecg{\theta})$.
Here, the parameters $\vecg{\varphi}$ describe the form of this likelihood, and are found by iteratively fitting to simulations.
After fitting, the approximate likelihood $p_{\vecg{\varphi}}(\vec{x}\mid\vecg{\theta})$ can be used, in conjunction with MCMC methods, to obtain samples from the posterior without the need for additional simulations.
SNL is particularly well-suited for \emph{targeted inference}, where a model is trained to fit a specific observation $\vec{x}_{\mathrm{o}}$ with fewer simulations and a shorter training time.
This is because, after each iteration, the method requests additional simulations in the regions of parameter space where the value of $p(\vecg{\theta}\mid\vec{x}_{\mathrm{o}})$ is high and more accuracy is needed.

The primary goal of this paper is to show, for the first time, the viability of SNL methods for the search of MBHB GW signals in synthetic LISA data. In this sense, our study constitutes a proof of concept and, for this reason, we consider its application to synthetic data containing a single MBHB event.
The waveform model that we use for MBHBs is one of the families of compact binary coalescences developed in the context of data analysis of ground-based detectors, namely the IMRPhenomD model~\cite{Husa:2015iqa,Khan:2015jqa}.
Due to the large masses involved, MBHBs emit very strong GW signals, with SNRs that can surpass $1000$.
A comprehensive review of the astrophysics of MBHBs can be found in Sec.~2 of \cite{LISA:2022yao}.
The MBHB targeted has the same parameters as the first MBHB injected in the LDC2-b (codenamed \emph{Spritz}).
\emph{Spritz} focuses on the detection of isolated sources in the presence of glitches, gaps, and other Non-stationarities in the data, but we leave those for future work and focus on the viability of the method in the presence of stationary Gaussian noise.
Then, the investigations presented here are closer in nature to the first LDC (codenamed \emph{Radler}), but with using Keplerian orbits for the LISA constellation and second-generation Time-Delay Interferometry (TDI) of \emph{Spritz} included in our simulations.
Once the method is proven to work reliably with these conditions, we expect to be able to easily scale it to solve \emph{Spritz} in future work.

The fact that SNL has not yet been applied to GW parameter estimation is likely due to the fact that its viability is highly dependent on the dimensionality of $\vec{x}$. In this sense, we note that SNL gives us the freedom to do any transformation to $\vec{x}$ -- including the application of dimensionality reduction algorithms -- as long as it is done in an independent step to the inference itself. This work proposes two approaches: (i) linear compression using Principal Component Analysis (PCA), and (ii) nonlinear, neural network-powered dimensionality reduction using \emph{autoencoders}.

Finally, it is important to mention that our implementation of the SNL algorithm is quite modular and contains a number of ingredients whose characteristics and integration in the global method have a considerable room for improvement. The list of tunable components includes: choice of waveform model and data representation, choice of posterior sampler, choice of dimensionality reduction algorithm, and the choice of neural network architecture (or architectures, if using autoencoders). This paper constitutes a proof of concept and future research will address each of these elements to improve the global performance and reliability of the algorithm.

The structure of this article is as follows: Sec.~\ref{sec:data-analysis-basics} briefly reviews some basic concepts of GW data analysis used in this paper.
Sec.~\ref{sec:nle-base} begins with a broad overview of SBI before giving a focused explanation of SNL, the main data analysis algorithm used in this article.
Before being able to apply it, a simulator that produces data in a form appropriate for SNL is required, so Sec.~\ref{sec:sim-pipeline} details the basic simulation pipeline that we have developed.
In anticipation to the analysis of GW sources in data containing glitches and gaps in the future, Sec.~\ref{sec:modified-pipeline} describes some modifications to this pipeline which were also tested here.
In Sec.~\ref{sec:results}, we present the experiments we performed and the results obtained from them.
Finally, Sec.~\ref{sec:discussion} summarizes our findings and the conclusions drawn from them, and outlines the next steps in this line of research.
As for the appendices, Appendix~\ref{sec:intervals-and-downsampling} details how the time interval and the sampling cadence for the simulations of Sec.~\ref{sec:waveforms-and-noise} are chosen.
Appendix~\ref{sec:noise-psd} contains the equations for the LISA noise PSD model that is used in this work.
Appendix~\ref{sec:ae-architecture} gives the hyperparameters for the autoencoder architecture used in Sec.~\ref{sec:autoencoders} and throughout our experiments in Sec.~\ref{sec:results}.

\section{\label{sec:data-analysis-basics}Likelihood-based inference basics}
In this section, we briefly review the basics of GW data analysis, which nowadays mostly relies on Bayesian sampling methods, with particular focus in the case of LISA.
Given some data $\vec{x}$, we are ultimately interested in finding the \emph{posterior distribution} $p(\vecg{\theta}\mid\vec{x})$ over the GW signal parameters $\vecg{\theta}$.
The posterior can be obtained using Bayes' theorem:
\begin{equation} \label{eq:bayes}
    p(\vecg{\theta} \mid \vec{x}) = \frac{p(\vec{x}\mid\vecg{\theta}) \, p(\vecg{\theta})}{p(\vec{x})} \,,
\end{equation}
where $p(\vec{x}\mid\vecg{\theta})$ is the \emph{likelihood} of obtaining the data realization $\vec{x}$ containing a GW signal with parameters $\vecg{\theta}$; $p(\vecg{\theta})$ is the \emph{prior} distribution, describing our knowledge of the signal parameters before observing the measurement; finally, $p(\vec{x})$ is the \emph{evidence} of the data.

In GW data analysis, it is assumed that the data has the additive functional form:
\begin{equation}
    \vec{x}(t) = \vec{h}^{}_{\vecg{\theta}}(t) + \vec{n}(t)\,,
\end{equation}
where $\vec{h}_{\vecg{\theta}}(t)$ is the part of the data corresponding to a GW signal characterized by a set of physical parameters $\vecg{\theta}$ and can be deterministically modelled\footnote{This is not the case when we consider the presence of stochastic GW backgrounds, but we ignore them in this work.}.
On the other hand, $\vec{n}(t)$ denotes the \emph{noise}, coming from instrumental and environmental effects.
Assuming that the noise is stationary and Gaussian, the likelihood function can be defined~\cite{Finn:1992wt, Cutler:1994ys}, up to a normalization factor, as
\begin{equation}
    p(\vec{x}\mid\vecg{\theta}) \propto \exp\left\{- \frac{(\vec{x} - \vec{h}_{\vecg{\theta}} \mid \vec{x} - \vec{h}_{\vecg{\theta}})}{2}\right\} \,,
    \label{eq:likelihood}
\end{equation}
where $(\vec{a} \mid \vec{b})$ is the noise-weighted inner product, defined as
\begin{equation}
    (\vec{a} \mid \vec{b}) = 4 \Re \sum_{i,j} \int_0^\infty \conj{\fourier{a}_i}(f) [S_n(f)^{-1}]^{ij} \fourier{b}_j(f) \, \dee f \,.
    \label{eq:inner-product}
\end{equation}
The sum in this inner product is over all the detector channels, and $[S_n(f)]^{ij}$ denotes the noise power spectral density matrix correlated between detector channels $i$ and $j$.
In the case of a network of ground-based detectors situated at different locations, it is common to assume that the noises at each detector are uncorrelated\footnote{For the Einstein Telescope, where multiple detectors are planned to be in the same location, this is no longer the case.}, and therefore $[S_n(f)]^{ij} = 0$ for $i \neq j$.
In LISA, however, the noises at each spacecraft are correlated, which translates into correlations between the TDI observables $X$, $Y$, and $Z$. Instead, we can construct TDI combinations $(A, E, T)$ that are uncorrelated, which simplifies the likelihood computation.
Furthermore, the sensitivity to GWs of the TDI channel $T$ is heavily suppressed at low frequencies~\cite{Prince:2002hp,Tinto:2020fcc}. 
In the case of MBHB signals, it is safe to ignore this channel, as we can see from the frequency range covered by MBHBs that is discussed inside Appendix~\ref{sec:intervals-and-downsampling}, in particular in Figure~\ref{fig:sampling_rate_choice}.

The inner product in Eq.~\eqref{eq:inner-product} is also used to define quantities useful for GW data analysis, such as the \emph{signal-to-noise ratio} (SNR)
\begin{equation} \label{eq:snr}
    \mathrm{SNR}[\vec{h}_{\vecg{\theta}}] = (\vec{\vec{h}_{\vecg{\theta}}} \mid \vec{\vec{h}_{\vecg{\theta}}})^{1/2} \,,
\end{equation}
and the \emph{overlap} between a waveform $\vec{h}_{\vecg{\theta}}$ and a measured signal $\vec{s}$,
\begin{equation} \label{eq:overlap}
    \mathcal{O}[\vec{h}_{\vecg{\theta}}\mid\vec{s}] = \frac{(\vec{h}_{\vecg{\theta}}\mid\vec{s})}{\sqrt{(\vec{h}_{\vecg{\theta}}\mid\vec{h}_{\vecg{\theta}})}} \,.
\end{equation}
Given the likelihood function defined in Eq.~\eqref{eq:likelihood}, one can sample from the posterior distribution using Eq.~\eqref{eq:bayes}.
The evidence, $p(\vec{x})$, is usually numerically intractable and hence it is usually ignored and normalization of the posterior probability distribution is performed after sampling\footnote{To be more specific, MCMC sampling obtains samples of the (unnormalized) posterior, and by normalizing them one can also obtain an estimate of the evidence. The main goal of nested sampling is to compute the evidence, but one also ends with a set of posterior samples as a byproduct.}.
The evidence becomes relevant when the number of sources present in the data is unknown, or when comparing between different waveform models, but those scenarios lie outside of the scope of this work.

As for the sampling algorithm itself, it is necessary that it is robust to the appearance of degeneracies and multimodalities in the posterior.
As mentioned before, successful submissions in past iterations of the LDCs have included methods based on nested sampling and parallel-tempered MCMC.
This work will make use of the \texttt{Eryn} sampler~\cite{Karnesis:2023ras,katzMikekatz04ErynFirst2023}, an ensemble-based MCMC sampler with capabilities for both parallel tempering and Reversible Jump MCMC, developed to fulfill the requirements for LISA data analysis.

Parallel tempering, also known as \emph{replica exchange} MCMC, consists of running multiple MCMC chains in parallel, each sampling the posterior weighted by an assigned inverse temperature $\beta$:
\begin{equation}
    p(\vecg{\theta}\mid\vec{x}; \beta) \propto p(\vec{x} \mid \vecg{\theta})^{\beta} \, p(\vecg{\theta}) \,.
\end{equation}
The chains with higher temperatures (lower $\beta$) thus give higher priority to the prior, sampling a more spread out posterior.
There is a mechanism by which chains with different temperatures can exchange their state.
Therefore, if a high-temperature chain finds a region of high posterior density, it will exchange its state with a lower-temperature chain, which is able to explore it more thoroughly.

\section{\label{sec:nle-base}Neural Likelihood Estimation}
In this section, we introduce the basic concepts of Neural Likelihood Estimation, which can be found in the literature of Machine Learning methods. We introduce these concepts in a generic way, mostly independent of any physical problem, and we then proceed from a broad discussion to specific details.

\subsection{\label{sec:sbi}Simulation-based inference}
Given a forward simulator $\vec{x} = \vec{x}(\vecg{\theta})$, SBI aims at obtaining the posterior distribution $p(\vecg{\theta} \mid \vec{x})$ of the parameters $\vecg{\theta}$ for observed data $\vec{x}_\mathrm{o}$.
The way in which this is achieved is by replacing some part of the posterior (see below), as given by Bayes' theorem in Eq.~\eqref{eq:bayes}, by a flexible estimate, for which evaluation and/or sampling is fast.
The set of parameters $\vecg{\varphi}$ can be tuned from a set of simulations and the set of parameters used to generate them,  $\{\vec{x}(\vecg{\theta}), \vecg{\theta}\}$.
There are three main approaches to SBI:
\begin{description}[leftmargin=*]
\item[Neural Posterior Estimation (NPE)] This is the most straightforward method (see~\cite{Papamakarios:2016ctj, 2017arXiv171101861L, 2019arXiv190507488G}).
        The posterior $p(\vecg{\theta} \mid \vec{x})$ is replaced by an estimate $p_{\vecg{\varphi}}(\vecg{\theta} \mid \vec{x})$.
        Inference for some observed data $\vec{x}_{\mathrm{o}}$ then simply consists of drawing samples from $p_{\vecg{\varphi}}(\vecg{\theta} \mid \vec{x}_{\mathrm{o}})$, which should be computationally fast by construction.
        This is the optimal approach if we want to perform \emph{amortized inference} -- if we have multiple separate observations, we can substitute each of them in our already-trained estimate and quickly obtain their respective posteriors without any need to further retrain our model.
        NPE has been applied to ground-based GW data analysis by the DINGO program \cite{Dax:2021tsq, Dax:2021myb, Dax:2022pxd}, with resounding success.
        Ref.~\cite{Korsakova:2024sut} also applies NPE techniques to LISA data containing signals from galactic binaries.
        However, the posterior estimate is trained on simulations drawn from a certain $p(\vecg{\theta})$, and changing this -- i.e., adapting to new priors  -- requires either the implementation of one of a few possible workarounds (see, e.g.~\cite{Papamakarios:2016ctj, 2017arXiv171101861L, 2019arXiv190507488G}), which may in turn impose some restrictions on the form of the estimate or priors, or the retraining of the model from scratch.

\item[Neural Likelihood Estimation (NLE)] Modelling the likelihood as $p_{\vecg{\varphi}}(\vec{x} \mid \vecg{\theta})$ (see~\cite{Papamakarios:2018zoy,2018arXiv180509294L,Alsing:2019xrx}) avoids the dependence on the prior present in NPE.
        Given our data $\vec{x}_{\mathrm{o}}$ and some prior $p(\vecg{\theta)}$, we can then quickly evaluate the unnormalized posterior as $p(\vecg{\theta} \mid \vec{x}_{\mathrm{o}}) \propto p(\vec{x}_{\mathrm{o}} \mid \vecg{\theta}) \, p(\vecg{\theta})$.
        However, it is necessary to draw samples from the Bayesian posterior, for instance via MCMC algorithms, making amortized inference less efficient than in NPE, but still much faster than likelihood-based methods.
        Conversely, since $p_{\vecg{\varphi}}(\vec{x} \mid \vecg{\theta})$ is prior-independent\footnote{But taking into account that $p_{\vecg{\varphi}}(\vec{x} \mid \vecg{\theta})$ is only well-defined within the region of parameter space used in training the model.}, NLE allows for active learning for a particular data set $\vec{x}_{\mathrm{o}}$ by actively selecting the values of $\vecg{\theta}$ that are simulated while the model is being trained.
        This should make it feasible to perform shorter \emph{targeted} training runs.
        The main drawback of this method is that the likelihood function has the same dimensionality as our data -- in other words, the larger the size of $\vec{x}$, the more complex and computationally expensive our model will need to be in order to achieve a similar performance, making dimensionality reduction of $\vec{x}$ a critical task.
        To the best of our knowledge, NLE has yet to be explored in GW data analysis for this very reason, and it is precisely the main subject that we have explored in this paper.

\item[Neural Ratio Estimation (NRE)] The aim is to estimate the likelihood-to-evidence ratio $r(\vec{x}, \vecg{\theta}) \equiv p(\vec{x} \mid \vecg{\theta})/p(\vec{x}) = p(\vec{x}, \vecg{\theta}) / p(\vec{x}) p(\vecg{\theta})$ (see \cite{2016arXiv161110242T,Hermans:2019ioj,2020arXiv200203712D,Miller:2021hys}).
        This approach effectively converts the density estimation problem into a much simpler classification one.
        Although sampling is still needed to obtain the posterior, NRE allows for easy marginalization over nuisance parameters. NRE has already been applied to GW data analysis (see~\cite{Bhardwaj:2023xph}).
\end{description}

\subsection{Normalizing flows}
Both NPE and NLE rely on having a probability density function estimator -- that is, a family of functions $\{p_{\vecg{\varphi}}(\vec{x})\}$ that can flexibly take the form of probability distributions of $\vec{x}$, depending on the values of a set of parameters $\vecg{\varphi}$, to approximate a target distribution $p(\vec{x})$.
Neural networks are, by construction, parametric families of nonlinear functions, which can be arbitrarily complex depending on their architecture.
This makes the weights and biases that describe them a natural choice for $\vecg{\varphi}$, whose values can be determined by fitting using standard gradient descent techniques.
The problem thus becomes to construct an arbitrary $p_{\vecg{\varphi}}(\vec{x})$ from the outputs $\vec{y} = g_{\vecg{\varphi}}(\vec{x})$ of a neural network $g_{\vecg{\varphi}}$ which we are free to design -- a \emph{neural density estimator} (NDE).

The first NDE model proposed are Mixture Density Networks (MDNs) \cite{bishopMixtureDensityNetworks1994}.
These simply map the outputs of a neural network to the parameters that describe a mixture of simpler probability distributions -- usually, the weights, means and covariance matrix components of a mixture of Gaussian distributions.
MDNs are easy to implement, but their flexibility is limited by the number of components in the mixture -- which, in turn, dictates the number of outputs the neural network must have.
If the distribution we are targeting resembles the components (i.e. if it is close to a Gaussian), a few components already give satisfactory results.
Otherwise, the number of required components for a faithful estimate of the target distribution can quickly increase.

Normalizing flows \cite{Rezende:2015ocs,Kobyzev:2019ydm} reframe the problem of finding the distribution $p_{\vecg{\varphi}}(\vec{x})$ into one of converting it to a distribution $q(\vec{z})$ which is easier to evaluate and sample -- typically, a normal distribution with zero mean and unit variance.
This is done by finding a bijective transformation $\vec{x} = f(\vec{z})$ between the respective distributions sample spaces\footnote{The transformation $f(\vec{z})$ is going to be described by the network parameters $\vecg{\varphi}$, but we drop them from our notation for the sake of simplicity.}.
Sampling can then be directly performed on $q(\vec{z})$ and then fed through $f(\vec{z})$ to obtain samples of $p_{\vecg{\varphi}}(\vec{x})$.
To evaluate $p(\vec{x})$ for a specific value of $\vec{x}$, the inverse transformation $\vec{z} = f^{-1}(\vec{x})$ is done first, then $q(\vec{z})$ is evaluated.

For single-variable distributions $q(z)$ and $p_{\vecg{\varphi}}(x)$, the relationship between them is given by the change-of-variable formula:
\begin{equation}
    p_{\vecg{\varphi}}(x) = q(z) \left| \frac{\dee z}{\dee x} \right| = q(f^{-1}(x)) \left| \frac{\dee f^{-1}(x)}{\dee x} \right|\,,
\end{equation}
while for multivariate $p(\vec{z})$ and $q(\vec{x})$, the derivative becomes a Jacobian determinant
\begin{equation}
    p_{\vecg{\varphi}}(\vec{x}) = q(\vec{z}) \left| \det \mat{J}[f^{-1}(\vec{x})] \right| = \frac{q(\vec{z})}{\left| \det \mat{J}[f(\vec{z})] \right|}\,.
\end{equation}
For a generic transformation $f(\vec{x})$ that is both differentiable and invertible, computing the Jacobian determinant can be reduced to a numerical computation that scales like $\order(D^3)$ with the dimensionality $D$ of $\vec{x}$.
This makes it computationally unfeasible in practice.
However, if $f(\vec{x})$ is restricted to the subset of transformations that have a triangular Jacobian, the determinant becomes the product of its diagonal elements, which scales like $\order(D)$.
This greatly reduces the complexity of the transformations considered.
But, since the composition of two bijective differentiable functions is itself bijective and differentiable, complex transformations can be recovered by composing $k$ simple ones into a \emph{flow}
\begin{equation}
    f(\vec{z}) = \left[ f^{}_{k} \circ f^{}_{k-1} \circ \cdots \circ f^{}_2 \circ f^{}_1 \right] (\vec{z}) \,.
\end{equation}
Then, the change of variable formula becomes:
\begin{equation}
    p^{}_{\vecg{\varphi}}(\vec{x}) = q(\vec{z}) \prod_{i=1}^k \left| \det \mat{J}[f^{}_i(\vec{z}^{}_{i-1})] \right|^{-1}\,,
\end{equation}
with $\vec{z}_0 \equiv \vec{z}$.

\subsection{Masked Autoregressive Flows}
The problem has now been reduced to finding a family of bijective transformations $f_i(\vec{z})$ with a triangular Jacobian matrix.
The parameters that describe this family should be expressible as the output of a neural network.
There are several options in the literature \cite{Kobyzev:2019ydm}. In this work, we choose to implement Masked Autoregressive Flows (MAFs)~\cite{Papamakarios:2017tec}. These are described as an element-wise set of affine transformations:
\begin{equation}
    x^{}_j(\vec{z}) = \mu^{}_j + \sigma^{}_j z^{}_j \,,
\end{equation}
with $\mu_j = \mu_j(x_1,\ldots,x_{j-1})$, $\sigma_j = \sigma_j(x_1,\ldots,x_{j-1})$.
The value of $x_j$ thus depends only on the previous values of \vec{x} -- this is the so-called \emph{autoregressive property}.
It is straightforward to prove that the Jacobian of this transformation is by construction triangular, and its determinant is:
\begin{equation}
    \left|\det \mat{J}[f(\vec{z})]\right| = \prod^{}_i \sigma^{}_i \,.
\end{equation}
If the distribution $q(\vec{z})$ is chosen to be the standard normal distribution, then $z_j \sim \normal(0, \identity)$.
Scaling this by $\sigma_j$ and shifting by $\mu_j$ makes it so that $x_j \sim \normal(\mu_j, \sigma_j)$. Then, a single transformation is equivalent to modelling $p_{\vecg{\varphi}}(\vec{x})$ as
\begin{eqnarray}
    \label{eq:maf-likelihood}
    p^{}_{\vecg{\varphi}}(\vec{x}) & = & \prod_{j=1}^D p^{}_{\vecg{\varphi}}(x_j \mid x^{}_1, \ldots, x^{}_{j-1}) \nonumber \\[2mm]
    & = & p^{}_{\vecg{\varphi}}(x^{}_1) p^{}_{\vecg{\varphi}}(x^{}_2 \mid x^{}_1) p^{}_{\vecg{\varphi}}(x^{}_3 \mid x^{}_2, x^{}_1)\cdots \,,
\end{eqnarray}
\begin{equation} \label{eq:maf-likelihood-partial}
    p^{}_{\vecg{\varphi}}(x^{}_j \mid x^{}_1, \ldots, x^{}_{j-1}) = \normal(x^{}_j \mid \mu^{}_j, \sigma^{}_j) \,.
\end{equation}
More complex distributions can then be recovered by chaining several transformations, with the ordering of $x_i$ randomly permuted between them.

The functions $\mu_j(x_1,\ldots,x_{j-1})$ and $\sigma_j(x_1,\ldots,x_{j-1})$ for each transformation can be finally parameterized by the outputs of a neural network that takes $\vec{x}$ as input and gives the values of $\mu_j$ and $\alpha_j = \log \sigma_j$ in a single forward evaluation\footnote{The quantities $\alpha_j$ are used in place of the quantities $\sigma_j$ to automatically enforce the positivity of $\sigma_j$ without having to place additional restrictions on the network architecture.}.
The network architecture must be constrained so that the autoregressive property of the outputs is satisfied -- the output neurons corresponding to $\mu_j$ and $\alpha_j$ must not be connected to any of the inputs $x_j, x_{j+1},\ldots,x_D$.
In Ref.~\cite{Germain:2015yft}, these restrictions are enforced on fully-connected neural network architectures by cleverly constructing binary masks for each layer.
The resulting network architecture is called a Masked Autoencoder for Distribution Estimation (MADE).
Several MADEs can then be chained one after the other in order to form a MAF.
Between each MADE, the ordering of the data vector can be randomly permuted so that the dependency on the sequential order of the data is nullified.

In SBI, we are interested in estimating the likelihood conditional on the parameters $\vecg{\theta}$.
Conditioning can be added to MADEs by appending $\vecg{\theta}$ to the network input without any masking.
Then, training a conditional MAF consists in feeding it a set of $\{ \vec{x}_i, \vecg{\theta}_i \}$ pairs to obtain the parameters $\mu_j, \alpha_j$ from each MADE, from which the log-likelihood $\log p_{\vecg{\varphi}}(\vec{x}_i \mid \vecg{\theta}_i)$ can be computed by repeated application of Eqs.~\eqref{eq:maf-likelihood}-\eqref{eq:maf-likelihood-partial}.
Then, an optimization algorithm can be used to tweak the values of the weights $\vecg{\varphi}$ in order to minimize the loss function
\begin{equation}
    L = - \log p^{}_{\vecg{\varphi}}(\vec{x} \mid \vecg{\theta}) \,.
\end{equation}
In other words, our goal is to find the set of weights that maximize the estimated log-likelihood throughout the dataset.
This ensures that the likelihood estimate $p_{\vecg{\varphi}}(\vec{x} \mid \vecg{\theta})$ obtained after training closely mimics the real underlying likelihood $p(\vec{x} \mid \vecg{\theta})$ without ever having to explicitly evaluate it.

\subsection{\label{sec:snl}Sequential Neural Likelihood}
Out of the three SBI paradigms described in Sec.~\ref{sec:sbi}, this work focuses on likelihood estimation.
More specifically, we use Sequential Neural Likelihood (SNL) \cite{Papamakarios:2018zoy} to perform inference targeted to a specific data set $\vec{x}_{\mathrm{o}}$.
In brief, a dataset $\{ \vec{x}_i, \, \vecg{\theta}_i \mid i=1,\ldots,N_{\mathrm{sim}} \}$ is generated from the joint distribution $p(\vecg{\theta}, \vec{x}) = p(\vec{x}\mid\vecg{\theta}) \, p(\vecg{\theta})$ by first drawing $\vecg{\theta}$ from the prior distribution and obtaining the corresponding $\vec{x}$ through the simulator $\vec{x} = \vec{x}(\vecg{\theta})$.
This dataset is used to train our MAF and obtain a set of weights and biases $\vecg{\varphi}_1$ and their corresponding likelihood function $p_{\vecg{\varphi}_1}(\vec{x} \mid \vecg{\theta})$.
This would conclude single-round NLE.
SNL adds to this by iterating this analysis over multiple rounds.
We can draw $N_{\mathrm{sim}}$ new samples of $\vecg{\theta}$ from the (approximate) target posterior $p_{\vecg{\varphi}_1}(\vecg{\theta}\mid\vec{x}_{\mathrm{o}}) \propto p_{\vecg{\varphi}_1}(\vec{x}_{\mathrm{o}} \mid \vecg{\theta}) \, p(\vecg{\theta})$ using MCMC.
Those samples, and their corresponding simulated $\vec{x}$, are then appended to the dataset $\{ \vec{x}_i, \, \vecg{\theta}_i \}$, and training is resumed to update $\vecg{\varphi}_1$ into $\vecg{\varphi}_2$ and obtain $p_{\vecg{\varphi}_2}(\vec{x} \mid \vecg{\theta})$.
This process of sampling, simulation, and retraining is repeated until convergence is reached, or a computational budget (e.g. number of simulations or training time) is met.

In this way, SNL quickly finds the regions of parameter space with high posterior density, where more accuracy is desired, and obtains more training data there.
Because the training data from earlier rounds is not discarded, the simulations proposed in later rounds do not bias the likelihood estimate.
However, a model trained to target a specific realization of $\vec{x}_{\mathrm{o}}$ is not expected to perform as well on a different $\vec{x}_{\mathrm{o}}$, and it will perform poorly if its underlying parameters are outside the support of the initial prior $p(\vecg{\theta})$.
This drawback is partially offset by having much faster training with fewer overall simulations.
Targeted inference can considered to be a better option when high-accuracy simulations become expensive to run and few distinct observations are made.

Once a full round of training is done, we have a likelihood estimate $p_{\vecg{\varphi}}(\vec{x} \mid \vecg{\theta})$ that can be quickly evaluated without performing any additional simulations.
We are ultimately interested in obtaining samples of the posterior given some observed data $\vec{x}_{\mathrm{o}}$.
To obtain those, it is necessary to sample the posterior $p(\vecg{\theta} \mid \vec{x}_{\mathrm{o}}) \propto p_{\vecg{\varphi}}(\vec{x}_{\mathrm{o}} \mid \vecg{\theta}) \, p(\vecg{\theta})$.

\section{\label{sec:sim-pipeline}Simulation pipeline}
In order to train our model, large amounts of simulated MBHB coalescence data are needed.
We have created a data generation pipeline that can rapidly \emph{`mass-produce'} LDC-like datasets with arbitrary parameters.
Speed of simulation was prioritized over data fidelity and realism, in order to be able to rapidly develop, prototype and test our algorithms.
Our choice of waveform model is the frequency-domain IMRPhenomD \cite{Husa:2015iqa,Khan:2015jqa} model, which describes the $\ell= |m| = 2$  spin-weighted spherical-harmonic modes of the GW signal emitted during the inspiral, merger and ringdown phases of non-precessing (aligned-spin) binary BH coalescences. 
The intrinsic parameters of the IMRPhenomD waveform model are: (i) The masses of the BHs, $m_1$ and $m_2$; and (ii) the projected spins\footnote{IMRPhenomD only supports BH spins perpendicular to the orbital plane. The projected spin $\chi_i$ is the component of the spin projected on that axis. If the dimensionless spin ranges from $0$ to $1$, $\chi_i$ can range from $-1$ to $1$.} in the direction of the orbital angular momentum, $\chi_1$ and $\chi_2$.
There are some combinations of these parameters that have interest because the GW signal is specially sensitive to them.
In the case of masses we have the chirp mass
\begin{equation}\label{eq:chirp-mass}
\mathcal{M}_c = \frac{(m_1 m_2)^{3/5}}{(m_1 + m_2)^{1/5}} = \frac{q^{3/5}}{(1+q)^{1/5}}\, m_1 \,,  
\end{equation}
where $q=m_2/m_1$ is the mass ratio.  In the case of spins we have the effective spin parameter
\begin{equation}\label{eq:chi-effective-definition}
\chi_\mathrm{eff} = \frac{m_1 \chi_1 + m_2 \chi_2}{M} = \frac{\chi_1 + q\chi_2}{1+q}\,,   
\end{equation}
where $M=m_1+m_2$ is the total mass of the binary BH.
On the other hand, the extrinsic parameters of the IMRPhenomD model are: (i) the luminosity distance, $D_L$; (ii) the longitude and latitude of the binary BH sky location in the LISA frame~\cite{Marsat:2020rtl}, $\lambda_L$ and $\beta_L$ respectively; (iii) the coalescence time, $t_c$; (iv) the GW phase at coalescence time, $\phi_c$; (v) the inclination angle between the orbital angular momentum and the line of sight, $\iota$; and (vi) the GW polarization angle, $\psi_L$.
In total, the waveform model has 11 parameters (see Table~\ref{tab:parameters-priors-truths} for the values of these parameters assumed in this work and the associated prior distributions).

The preprocessing of the data makes several assumptions --- detailed in what follows --- that, for the sake of speed, may not be completely realistic.
Because of this, it is expected that our inference will not give satisfactory results when shown real data, but in the same way as other methods that use the same type of waveform model.
Nevertheless, it is always possible to replace our simulation pipeline with a more realistic one, without changing the overall scheme of the method, and train using data generated from it to obtain more robust results.
This is going to be part of future improvements of the method.
A summary of the procedure is given in Algorithm~\ref{alg:data-pipeline}, with more details given in the following subsections.

\begin{figure}
    \begin{algorithm}[H]
        \caption{\label{alg:data-pipeline}Fast data generation pipeline}
        \begin{algorithmic}
            \Require $\vecg{\theta} \sim p(\vecg{\theta})$ (or $p(\vecg{\theta} \mid \vec{x}_{\mathrm{o}})$)
            \State $(\fourier{A}(f), \fourier{E}(f), \fourier{T}(f)) \gets \Call{lisabeta}{\vecg{\theta}}$
            \State discard $\fourier{T}(f)$
            \For{$H \gets \left\{A, E\right\}$} \textbf{in parallel}
            \State $\fourier{H}(f) \gets \fourier{H}(f) \, \sqrt{\Delta t}/\sqrt{S_H(f)}$ \Comment{whitening}
            \State $H(t) \gets \Call{IFFT}{\fourier{H}(f)} / \Delta t$
            \If{noisy}
            \State sample $n(t) \sim \normal(0, \identity)$
            \State $H(t) \gets H(t) + n(t)$
            \EndIf
            \EndFor
            \State $\vec{d} \gets (A(t), E(t))$ \Comment{concatenate channels}
            \If{PCA} \Comment{section~\ref{sec:pca}}
            \State $\vec{x} \gets \Call{PCA}{\vec{d}}$ \Comment{already pre-fit PCA}
            \ElsIf{autoencoder} \Comment{section~\ref{sec:modified-pipeline}}
            \State $\vec{d} \gets (\vec{d} - \vecg{\mu}_s)/\vecg{\sigma}_s$ \Comment{scaling}
            \State $\vec{x} \gets E_{\vecg{\psi}}(\vec{d})$ \Comment{from pre-trained autoencoder}
            \Else $\ \vec{x} \gets \vec{d}$ \Comment{no dimensionality reduction}
            \EndIf
            \State \textbf{return} $\vec{x}$
        \end{algorithmic}
    \end{algorithm}
\end{figure}

\subsection{Choice of initial priors}

In a first step, values for the MBHB parameters $\vecg{\theta}$ need to be selected.
This is done by sampling from a prior $p(\vecg{\theta})$.
In the first round of SNL, the prior can be configured depending on our initial assumptions.
The settings used are shown in Table~\ref{tab:parameters-priors-truths}.
In later rounds, this prior is also used in the MCMC posterior sampling step.
Generally, $p(\vecg{\theta})$ denotes the broadest region of support for our algorithm.

\begin{table}
    \centering
    \begin{tabular}{cccl} 
        \hline
        Notation           & Parameter description                                                                                                                                                                                                                                                          & True value                                                                                             & Prior                                               \\
        \hline
        \midrule
        $\lambda^{}_L$     & Longitude (\unit{\radian})                                                                                                                                                                                                                                                     & 0.7222569995679988                                                                                     & $\mathcal{U}(\lambda^{}_L; 0, 2\pi)$                \\
        $\beta_L$          & Latitude  (\unit{\radian})                                                                                                                                                                                                                                                     & -0.6361052210272687                                                                                    & $\mathcal{U}(\sin \beta^{}_L; -1, 1)$               \\
        $\chi^{}_1$        & Projected spin of BH 1 & 0.26863190922667673                                                                                    & $\mathcal{U}(\chi^{}_1; -1, 1)$                     \\
        $\chi^{}_2$        & Projected spin of BH 2                                                                                                                                                                                                                                                         & -0.4215109787709388                                                                                    & $\mathcal{U}(\chi^{}_2; -1, 1)$                     \\
        $\Delta t^{}_c$    & Coalescence time deviation (\unit{\second})                                                                                                                                                                                                                                    & 0.0\footnote{The absolute coalescence time $t_c$ in the LISA frame has value $11526688.585926516\;$s.} & $\mathcal{U}(\Delta t^{}_c; \num{-e3}, \num{e3})$   \\
        $\phi^{}_c$        & Phase at coalescence (\unit{\radian})                                                                                                                                                                                                                                          & 1.2201968860015653                                                                                     & $\mathcal{U}(\phi^{}_c; 0, 2\pi)$                   \\
        $\iota$            & Inclination angle (\unit{\radian})                                                                                                                                                                                                                                             & 2.2517895222056112                                                                                     & $\mathcal{U}(\cos \iota; -1, 1)$                    \\
        $\psi^{}_L$        & Polarization angle (\unit{\radian})                                                                                                                                                                                                                                            & 2.4281809899277578                                                                                     & $\mathcal{U}(\psi^{}_L; 0, 2\pi)$                   \\
        $D^{}_L$           & Luminosity distance (\unit{\mega\parsec})                                                                                                                                                                                                                                      & 13470.983558972537                                                                                     & $\mathcal{U}(D^{}_L; \num{3000}, \num{e5})$         \\
        $\mathcal{M}^{}_c$ & Chirp mass (\unit{\solarmass})                                                                                                                                                                                                                                                 & 772462.857152831                                                                                       & $\mathcal{U}(\mathcal{M}^{}_c; \num{e5}, \num{e7})$ \\
        $q$                & Mass ratio\footnote{We use the convention of $m_2 \leq m_1$, so $q \leq 1$.} $m_2/m_1$                                                                                                                                                                                         & 0.46285493073209505                                                                                    & $\mathcal{U}(q; \num{0.125}, \num{1})$              \\
        \hline
    \end{tabular}
    \caption{\label{tab:parameters-priors-truths}Definition, true values and initial priors used for the parameters $\vecg{\theta}$ describing the MBHB system being analyzed. $\mathcal{U}(\theta; a, b)$ denotes a prior uniform in $\theta$ ranging from $a$ to $b$. The longitude, latitude and polarization angle refer to the LISA frame.}
\end{table}

The main assumption made by our prior choice is that an initial estimate of the time of coalescence, $\bar{t}_c$, has been found through an initial search algorithm.
This allows us to define the error in this estimate $\Delta t_c = t_c - \bar{t}_c$ as the parameter that our algorithm is attempting to find.
The true value of $t_c$ in seconds may be large, but the posteriors are expected to be extremely narrow in comparison, which may give rise to floating point precision errors.
This parameter redefinition circumvents those errors.
As long as the error is within the prior for $\Delta t_c$, the estimate $\bar{t}_c$ does not need to be very accurate -- for our purposes, it could be even found by eye.
However, we do not concern ourselves with the search, and set an ideal $\bar{t}_c = t_c$.

\subsection{\label{sec:waveforms-and-noise}Waveform generation and noise}

For every set of MBHB parameters $\vecg{\theta}$, a MBHB simulation is then run using the \texttt{lisabeta} code (see \cite{Marsat:2018oam,Marsat:2020rtl}).
The waveform model supplied by this code is an accelerated version of the \texttt{IMRPhenomD} model~\cite{Husa:2015iqa,Khan:2015jqa}, using the formalism described in Ref.~\cite{Marsat:2018oam} to apply the LISA instrumental response in the frequency domain and directly output TDI observables.
The comparatively slower time-domain calculation is thus avoided.
The outputs of the code are the second-generation TDI  $(A, E, T)$ channels~\cite{Tinto:2022zmf, Tinto:2020fcc} in frequency domain -- though the $T$ channel was promptly discarded, as the amount of GW information it carries is suppressed at low frequencies~\cite{Prince:2002hp,Tinto:2020fcc}.
In connection to this, in Appendix~\ref{sec:intervals-and-downsampling} we provide information about the maximum frequency of MBHB events (see also Fig.~\ref{fig:sampling_rate_choice}), which justifies this approximation.
Finally, Keplerian LISA orbits are used in the computation of the waveform response.

In order to reduce simulation time and data dimensionality, the simulations are run in a restricted time range and at a lower sample rate than the LDCs.
The start and end-times of the simulation\footnote{At this stage of the pipeline, the TDI data is in the frequency domain. The start and end times determine two quantities: the time shift of the start of the waveform introduces a global phase factor of $e^{-i 2 \pi f (\bar{t}_c - t_b)}$, and the frequency resolution depends on the total observation duration by $\Delta f = (t_a + t_b)^{-1}$.} were fixed to a restricted range around the true coalescence time of the target -- from $t_b = \qty{24}{\hour}$ before, to $t_a = \qty{6}{\hour}$ after.
The sampling cadence is set to $\Delta t = \qty{15}{\second}$.
The reasoning behind these choices can be found in Appendix~\ref{sec:intervals-and-downsampling}.

Afterwards, instrumental noise can be added or not, depending on the type study to be done with the data.
In order to speed up the simulation, this work assumes that the noise is stationary and Gaussian, and that it follows \emph{a priori} known PSDs $S_n^{A,E}(f)$, which are introduced and described in Appendix~\ref{sec:noise-psd}.
We `\emph{whiten}' our frequency-domain noise-free TDI waveforms to have them in a form such that, if there is noise, it would be white and of unit amplitude.
After careful consideration of the normalization factors involved, the re-scaling needed to perform this whitening is
\begin{equation} \label{eq:whitening-transform}
    \fourier{A}(f) \rightarrow \sqrt{\frac{\Delta t}{S_n^A(f)}} \fourier{A}(f)\,,
\end{equation}
and the equivalent one for $\fourier{E}(f)$.
After this rescaling, one can then obtain a noisy time series under our assumptions by simply performing an Inverse Fast Fourier Transform (IFFT) and adding the noise as sampled from a standard normal distribution.
This completely bypasses the need to simulate the LISA instrument, greatly accelerating the production of large datasets.
Removing the assumptions on the noise is not expected to require significant changes to the inference algorithm, but realistic noise would slow down simulations and hence, it is left to future work.

\subsection{Dimensionality reduction: Principal Component Analysis \label{sec:pca}}
At this stage, our data consists of two channels (two time series), with 7200 data points each, which we can concatenate into a single 14400-dimensional vector $\vec{d} = (A(t), E(t))$.
Our chosen inference algorithm constrains our MADEs so that the number of inputs equals the dimensionality of the input $\vec{x}$ plus the number of parameters (11 in our case), and the number of MADE outputs will be twice the dimensionality of $\vec{x}$.
Using the original $14400$-dimensional data set, $\vec{x} = \vec{d}$, would require large neural networks that are computationally expensive to train and prone to overfitting.
The large number of inputs dedicated to $\vec{x}$ compared to $\vecg{\theta}$ may decrease the effect of $\vecg{\theta}$ during training, further slowing it down.
Therefore, an additional dimensionality reduction step is required to go from $\vec{d}$ into some compressed summaries $\vec{x}$.

NPE-based approaches such as DINGO~\cite{Dax:2021tsq} do not have as many problems with the dimensionality of $\vec{x}$, because they are estimating the probability of $\vecg{\theta}$ conditioned on $\vec{x}$.
This allows them to simultaneously train an embedding network that finds the optimal compression (down to a more manageable 128 dimensions) at the same time as they train their normalizing flows.
However, in the case of NLE, an embedding neural network on $\vec{x}$ cannot be trained at the same time as the inference networks.
Attempting to find a lower-dimensional form of $\vec{x}$ at the same time as one is training $p_{\vecg{\varphi}}(\vec{x}\mid\vecg{\theta})$, which is the probability to obtain the data, will quickly result in a compressed $\vec{x}$ such that the corresponding normalizing flow needs only be the identity mapping, completely disregarding any dependence on $\vecg{\theta}$.
Nevertheless, it remains possible to use dimensionality reduction algorithms to produce a compressed $\vec{x}$, as long as any required training is performed in a separate step.

The most basic strategy to compress the data used in this work is Principal Component Analysis (PCA) (see, e.g., \cite{jolliffePrincipalComponentAnalysis2002}).
The procedure is as follows: first, a dataset of 50000 waveforms, with parameters drawn from our prior distributions (see Table~\ref{tab:parameters-priors-truths}), is produced.
The elements of this dataset are stacked into a $50000\times 14400$ matrix onto which PCA is performed.
From this, we obtain a representation of this dataset in terms of the 14400 principal components derived from it.

Next, we need to evaluate how many of these principal components are needed.
The standard way of doing this is to look at the \emph{explained variance} ratio, which is a proxy of the fraction of the total information each principal component holds.
Thus, by looking at the cumulative explained variance ratio, we have a measure of the amount of information we preserve if we represent our data as a smaller number of these components.

\begin{figure}
    \centering
    \includegraphics[width=\columnwidth]{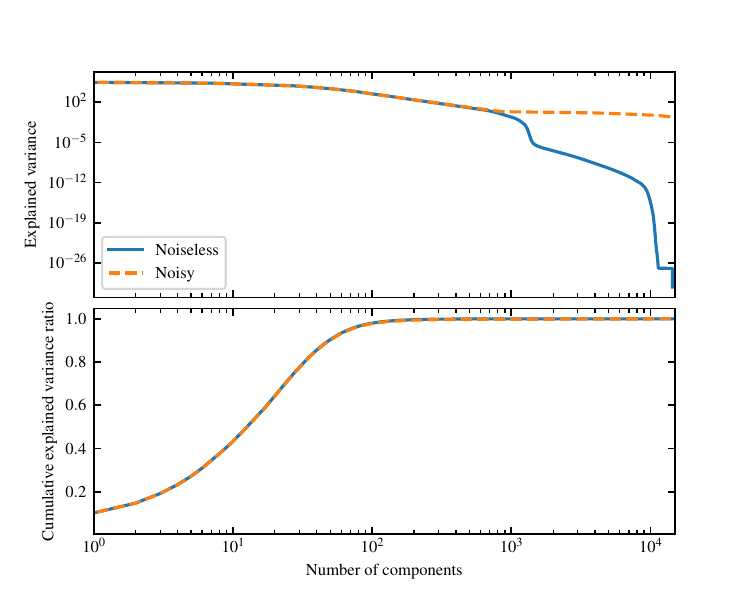}
    \caption{\label{fig:pca-noscaling}Top: Explained variance of each principal component when performing PCA on noise-free (solid) and noisy (dashed) data. Bottom: Cumulative explained variance ratio when truncating PCA to the number of components denoted in the $x$ axis.}
\end{figure}

The top panel of Fig.~\ref{fig:pca-noscaling} shows the explained variance of each principal component when performing PCA on both noise-free and noisy data.
In the case noise-free data, the explained variance begins to decrease notably after a few hundred components, falling off drastically after the first 1000.
The noisy data has a more gradual decrease of the component values, since the higher components are dedicated to describing the noise itself.
The bottom panel of Fig.~\ref{fig:pca-noscaling} shows the cumulative ratio.
In both, noisy and noise-free data sets, this ratio increases rapidly until it converges near 1 -- a perfect reconstruction -- after only a few hundred components.
To preserve 99\% of the information, we find that we need at least 132 components for the clean waveforms, or 147 for noisy data.
These correspond to compression ratios of 109.09 and 97.96, respectively.
This is a significant improvement, and it already makes our method computationally feasible.

To have a consistent data size throughout our experiments, we choose to keep to 128 components throughout this work.
This preserves 98.93\% of information in the noise-free case, and 98.70\% when we add noise to the data.
With the expectation that 50000 simulations are a representative enough sample of our data, we store the appropriately truncated matrices obtained by the PCA procedure so that they can be reused on new simulations to quickly transform a 14400-dimensional data $\vec{d}$ to a 128-dimensional compressed data $\vec{x}$.

\section{\label{sec:modified-pipeline}Modifications to the simulation pipeline}
The implementation of a numerical code for the basic simulation pipeline described in Sec.~\ref{sec:sim-pipeline} has been done with speed of prototyping and iteration as the first priority.
Each transformation done to the data after the waveform simulation -- whitening, inverse FFT, addition of noise, and PCA -- is implemented in the form of a modular block.
In this way, it is easy to implement new reusable data transformation blocks, which can then be chained in any order to change the form of the data $\vec{x}$ being given to the parameter estimation algorithm.
We have performed various experiments, and in this section we describe the most noteworthy ones.

\subsection{\label{sec:scaling}Waveform scaling}
An incidental advantage of whitening the signal is that it is also rescaled so that the noise in the time series has amplitudes of order $\order(1)$, which Machine Learning models have less numerical difficulty with than with the typical values of GW strain of order $\order(10^{-20})$.
If our signals were expected to have lower or comparable amplitude to the noise -- as is the case of stellar-mass compact binary coalescences in ground-based detector data -- this rescaling would be optimal.
However, for the case of MBHBs this is not the case: for loud sources, although the signal and noise have comparable amplitudes during the inspiral and ringdown, the signal during the merger can have amplitudes over two orders of magnitude higher.
Additionally, MBHB GW signals with higher SNR tend to have shorter, louder mergers, making this disparity even more pronounced.
If one proceeds with the data as is, the dimensionality reduction algorithms that follow in the pipeline will focus on the merger -- meaning that most of the information loss will come from the inspiral and ringdown.
Forcing the merger to have a comparable amplitude to the inspiral and ringdown would, in principle, give each phase of the waveform equal footing.
For these reasons, we have studied how to rescale the data to achieve this effect.

The scaling procedure used in this work is relatively simple.
First, in a manner similar to Sec.~\ref{sec:pca}, a dataset of 250000 whitened time-domain simulations is produced by drawing samples from the prior.
From these, the mean $\vecg{\mu}_s$ and standard deviation $\vecg{\sigma}_s$ are computed at each time (across the 250000 realizations).
Then, scaling simply consists of the following transformation:
\begin{equation}
    \vec{d} \rightarrow \frac{\vec{d} - \vecg{\mu}_s}{\vecg{\sigma}_s}\,.
\end{equation}
The values of $\vecg{\mu}_s$ and $\vecg{\sigma}_s$ are then stored so that this rescaling can be performed on new data.
We find that scaling noise-free data independently would give too much importance to the inspiral and ringdown parts of the waveforms due to the small variances there.
Therefore, the same $\vecg{\mu}_s$ and $\vecg{\sigma}_s$ that were computed on noisy data have been used to scale noise-free data.

\subsection{\label{sec:autoencoders}Autoencoders}

\begin{figure}
    \centering
    \includegraphics[width=\columnwidth]{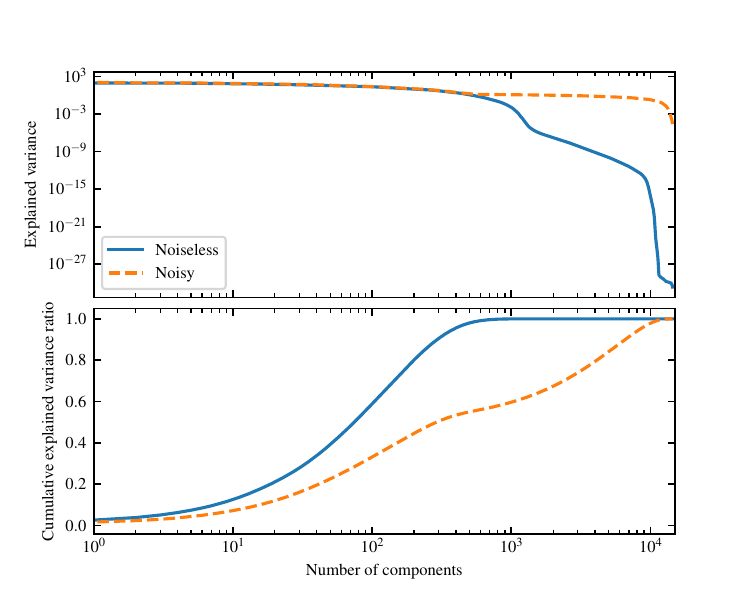}
    \caption{\label{fig:pca-scaling}Effect of reducing the number of components of PCA to the explained variance on data scaled with the procedure of Sec.~\ref{sec:scaling}. Top: Explained variance of each principal component when performing PCA on noise-free (solid) and noisy (dashed) data. Bottom: Cumulative explained variance ratio when truncating PCA to the number of components denoted in the $x$ axis.}
\end{figure}

After scaling the data, the next step in the pipeline would be to perform the dimensionality reduction.
The simplest way forward would be to use PCA on the scaled data, but we are going to show that this would result in severe loss of information.  The same procedure used to produce Fig.~\ref{fig:pca-noscaling} in Sec.~\ref{sec:pca} has been applied to scaled waveform data.
The explained variance of the resulting PCA is computed and shown on the top panel of Fig.~\ref{fig:pca-scaling}.
In the noise-free case, the explained variance only begins to decrease after 1000 components.  When we add noise, all components become almost equally important.

\begin{figure*}
    \centering
    \begin{subfigure}{0.45\textwidth}
        \centering
        \includegraphics[width=\linewidth]{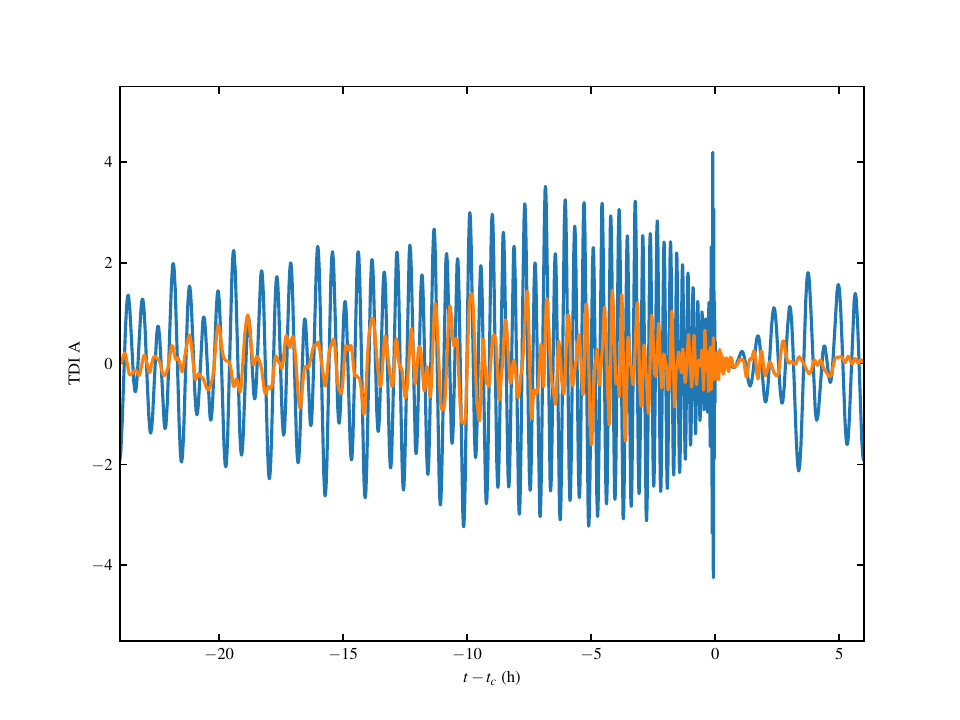}
        \caption{\label{fig:reconstruction-pca-noisefree}128-component PCA, noise-free}
    \end{subfigure}
    \begin{subfigure}{0.45\textwidth}
        \centering
        \includegraphics[width=\linewidth]{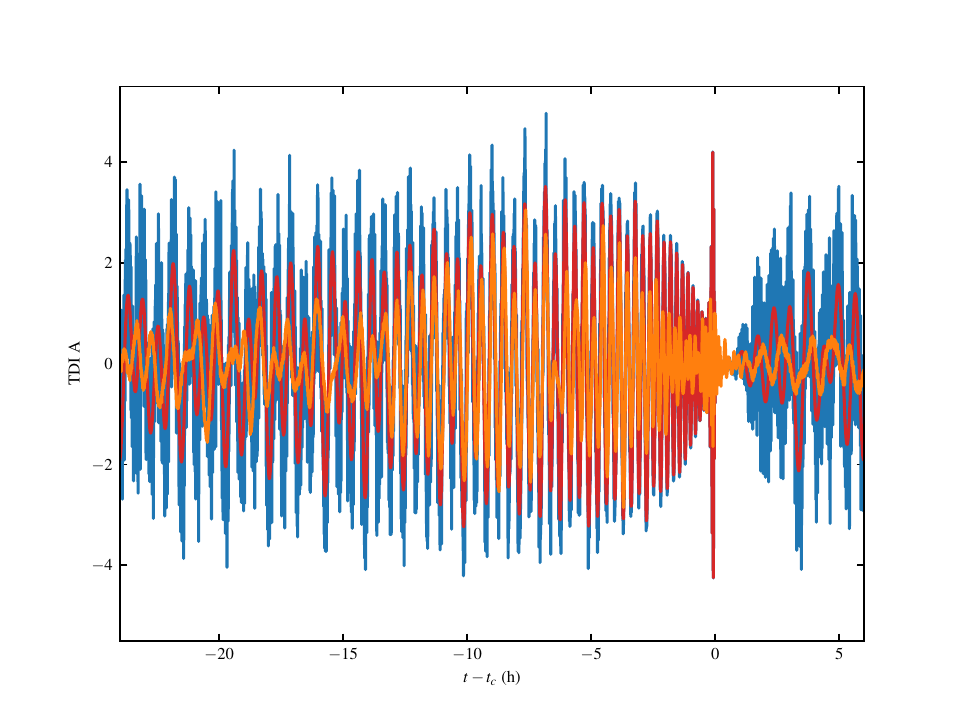}
        \caption{\label{fig:reconstruction-pca-noisy}128-component PCA, noisy}
    \end{subfigure}
    \begin{subfigure}{0.45\textwidth}
        \centering
        \includegraphics[width=\linewidth]{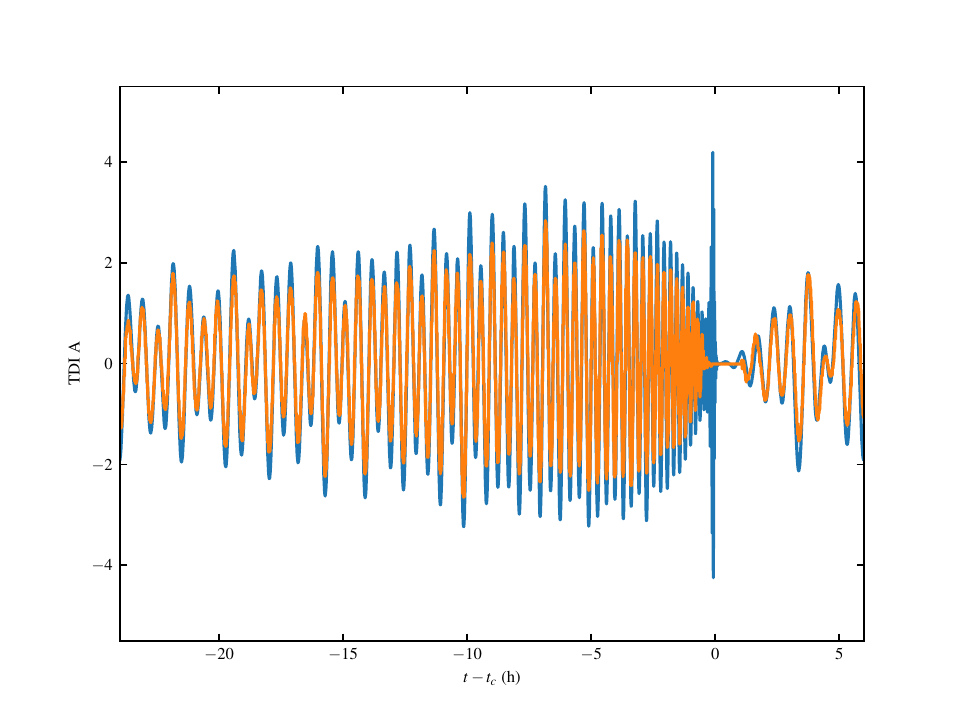}
        \caption{\label{fig:reconstruction-ae}Standard autoencoder, noise-free}
    \end{subfigure}
    \begin{subfigure}{0.45\textwidth}
        \centering
        \includegraphics[width=\linewidth]{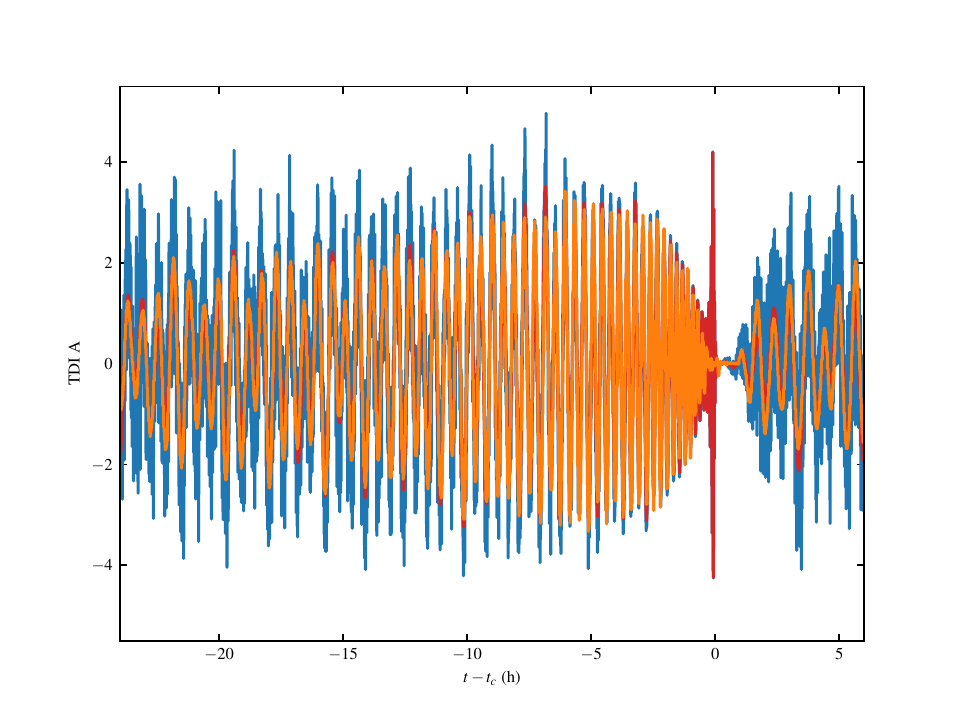}
        \caption{\label{fig:reconstrunction-dae}Denoising autoencoder, noisy}
    \end{subfigure}
    \caption{\label{fig:reconstructions}Reconstruction of the TDI A channel for PCA (above) and autoencoders (below). The left figures show performance on noise-free data: the original data is shown in blue, and the reconstruction after compression in orange. The right figures show noisy data: the colours are the same, but blue is now noisy data, with the added red showing the noise-free realization for reference.}
\end{figure*}

The cumulative ratio, shown in the bottom panel of Fig.~\ref{fig:pca-scaling}, gives clear indication that at least several hundred components would be needed to preserve most of the information in the noise-free case, or almost all of them when adding noise.
Indeed, in order to preserve $99$\% of the information, we found that at least $584$ components are needed for clean waveforms, while for noisy data a much larger figure of $10947$ components is required.
These are compression ratios of $24.66$ and $1.32$, respectively.
If we were to keep only the first $128$ components, we see that we would preserve less than $67$\% of the information in the noise-free case, and less than $38$\% in the presence of noise.
Figs.~\ref{fig:reconstruction-pca-noisefree}~and~\ref{fig:reconstruction-pca-noisy} show the scaled waveforms before\footnote{Fig.~\ref{fig:reconstructions} illustrates, in blue, the result of the scaling discussed in Sec.~\ref{sec:scaling}: the left subfigures show the noise-free scenario, and the right ones the scaled noisy data, with another replica of the noise-free data in red for reference.} and after the $128$-component PCA, in the noise-free and noisy cases, respectively.
It is important to remark that, while still not good enough, the PCA fit on noisy data results in a reconstruction that looks visually closer to the noise-free data than the PCA fit on noise-free data itself.
Still, the quality of the reconstruction is quite low, and it is for this reason that we conclude that linear dimensionality reduction via PCA is not powerful enough in the case of scaled data.

The next logical step is to use nonlinear dimensionality reduction algorithms. Here, we choose to test autoencoders \cite{Kramer:1991dqn,Hinton:2006tev}, since they are a conceptually simple but flexible technique.
An autoencoder consists of two function families with tunable parameters: an \emph{encoder} $E_{\vecg{\psi}}$ that projects the input data\footnote{While describing autoencoders, we change temporarily the notation: $\vec{d}$ becomes the high-dimensional vector $\vec{x}$, while $\vec{x}$ becomes the low-dimensional embedding $\vec{z}$.} $\vec{x}$ to a lower-dimensional latent space $\vec{z} = E_{\vecg{\psi}}(\vec{x})$, and a \emph{decoder} $D_{\vecg{\xi}}$ that takes a latent vector and produces another vector $\vec{y} = D_{\vecg{\xi}}(\vec{z})$ that belongs to the same vector space as $\vec{x}$.
If the parameters $(\vecg{\psi}, \vecg{\xi})$ are trained with the goal of achieving that $D_{\vecg{\xi}}(E_{\vecg{\psi}}(\vec{x})) \simeq \vec{x}$, the latent vector $\vec{z} = E_{\vecg{\psi}}(\vec{x})$ can be thought of as a lower-dimensional representation of $\vec{x}$.
Then, for a lower-dimensional $\vec{z}$, $\vec{y} = D_{\vecg{\xi}}(\vec{z})$ allows us to reconstruct something close to the original data vector $\vec{x}$.
The reconstruction error between $D_{\vecg{\xi}}(E_{\vecg{\psi}}(\vec{x}))$ and $\vec{x}$ -- in the form of a loss function -- becomes a proxy metric of how much information is lost in the compression from $\vec{x}$ to $\vec{z}$.
The typical loss function used is the mean-squared error (MSE) for a batch of $n_b$ examples:
\begin{equation}
    \textrm{MSE}(\vec{y}, \vec{x}) = \frac{1}{n^{}_b} \sum_{i=1}^{n_b} \lVert \vec{y}^{}_i - \vec{x}^{}_i \rVert_2^2 \,,
\end{equation}
with $\lVert \vec{x} \rVert_2 \equiv \sqrt{\vec{x} \cdot \vec{x}}$ denoting the $L^2$ norm (i.e. the Euclidean length).

In this work, the function families $E_{\vecg{\psi}}$ and $D_{\vecg{\xi}}$ are described by a convolutional neural network.
The autoencoder is designed to compress the 2-channel, $30$ hour long time series waveforms down to a vector of length $128$.
The details of the architecture of the autoencoder can be found in Appendix~\ref{sec:ae-architecture}.
Training is performed on a dataset of $200000$ noise-free, whitened and scaled waveforms drawn on the prior, until an early stopping criterion is met.
Performing a second, fine-tuning run on $200000$ additional simulations slightly improves the loss metric, but the improvement brought by further fine-tuning rounds is comparatively negligible.

As a visual test of the reconstruction capability of the autoencoder, our target waveform $\vec{x}_{\mathrm{o}}$ has been passed through the trained autoencoder.
Fig.~\ref{fig:reconstruction-ae} shows the reconstructed TDI A channel superimposed with the original.
While the recovery is not perfect, the order of magnitude and qualitative behaviour of the inspiral and ringdown parts of the waveform are fairly similar.
The merger, however, is completely missing from the reconstructed waveform.
This can be explained by the fact that, due to the high variance in frequencies near the merger, there are not enough convolutional filters to capture all of it.
Then, the network prioritizes learning the filters for the lower-frequency inspiral and ringdown, which explains a larger section of the time series.
Increasing the number of convolutional filters, especially in the first and last layers, seems to somewhat improve merger reconstruction, but also greatly increases the computational cost of training the autoencoder.
Considering that a compression factor of $112.5$ is being forced, and that the specific choice of architecture is highly unoptimized (as no hyperparameter tuning was performed over the number and configuration of layers), these are quite promising results.

In principle, noisy data could be problematic for data compression, but the flexibility of autoencoders gives us an additional way to combat the noise.
Instead of training to reconstruct the (noisy) input, the training is performed to minimize the MSE between the autoencoder output and the waveform without the noise.
In this sense, this \emph{denoising autoencoder} also learns to remove the noise from the data.
To ensure the network does not simply memorize the specific noise realizations in the examples, the noise can be quickly generated and added to the training waveforms before being fed to the autoencoder, effectively acting as a form of data augmentation.
Our experiments suggested that denoising autoencoders performed significantly better in noisy data than attempting to reconstruct the noisy signal.

The result of applying the denoising autoencoder to noisy data $\vec{x}_{\mathrm{o}}$ is shown in Fig.~\ref{fig:reconstrunction-dae}.
Both the noisy and noise-free waveforms are also shown.
Outside of the merger, the denoising autoencoder was able to reconstruct the noise-free waveform slightly better than the standard autoencoder.
At first, it may seem counterintuitive that the performance on noisy data is better than on noise-free data.
But this can be explained by the fact that we are explicitly guiding the autoencoder to learn to ignore the noise, which can help the autoencoder to more easily reconstruct the noise-free signal.

\begin{table}
    \centering
    \begin{tabular}{c|ccc}
        \hline
                              & \multicolumn{1}{r}{\textrm{Noise-free}} & \multicolumn{1}{r}{\textrm{Noisy}} & \multicolumn{1}{r}{\textrm{Noise}} \\
        \hline
        \midrule
        Random                & 3.0266                                  & 3.5556                             & 2.0007                             \\
        PCA-128               & 1.7647                                  & 2.2897                             & 1.0143                             \\
        Noisy PCA-128         & 1.2801                                  & 1.8087                             & 0.9785                             \\
        Autoencoder           & 0.1794                                  & 0.7867                             & 0.9919                             \\
        Denoising autoencoder & 0.1458                                  & 0.1398                             & 0.0005                             \\
        \hline
    \end{tabular}
    \caption{\label{tab:pca-mse}Reconstruction of the MSE on scaled data for the dimensionality reduction techniques discussed in this work -- lower is better.
    The three scenarios considered are: noise-free data, noisy data, and data consisting purely of noise.
    Each technique was applied on the scaled true $\vec{x}_{\mathrm{o}}$, followed by its inverse reconstruction.
    `Random' refers to sampling from a standard normal distribution.
    Since the denoising autoencoder always attempts to reconstruct the noise-free signal, that is what is given to its MSE -- in the case of pure noise, a zero signal is expected.}
\end{table}

For another comparison between PCA and autoencoders, Table~\ref{tab:pca-mse} shows the MSE resulting of applying each compression technique and its inverse on $\vec{x}_{\mathrm{o}}$, in the three cases where $\vec{x}_{\mathrm{o}}$ is noise-free, noisy and pure noise.
As a worst-case scenario to beat, they are compared with the MSE resulting from sampling a reconstructed data vector $\vec{x}_{\mathrm{o}}$ from a standard normal distribution -- which can be thought of as a completely lossy reconstruction process.
Unsurprisingly, for a specified compression factor, autoencoders outperform PCA in almost every scenario.
What is remarkable is that the denoising autoencoder seems to have performed slightly better than the standard one, even in the case where the data $\vec{x}_{\mathrm{o}}$ does have any noise -- despite the denoising autoencoder having never been shown purely noise-free data.
Indeed, the denoising autoencoder performs even slightly better than that when adding noise to the waveform.
This, along with the extremely low MSE when shown full noise, shows that the denoising autoencoder has learned how to effectively ignore the noise it was trained on.

In addition to the pipeline described in Sec.~\ref{sec:sim-pipeline}, we attempt to use scaled waveforms and autoencoders in the pipeline to test the performance of the data analysis algorithm described in Sec.~\ref{sec:nle-base}. In principle we do not expect the results to be particularly good -- after all, the autoencoder is failing to reconstruct the merger, where almost all of the SNR of the signal is concentrated. But, in any case, it serves as a proof of concept and a test of the robustness of our data analysis algorithm.

\section{\label{sec:results} Applications to synthetic data}

In order to test the feasibility of this algorithm, a synthetic data vector $\vec{x}_{\mathrm{o}}$ has been generated using the simulation pipeline summarized in Algorithm~\ref{alg:data-pipeline}, and described in detail in Sections~\ref{sec:sim-pipeline}~and~\ref{sec:modified-pipeline}.
The parameters of the source considered are the same ones as the first MBHB GW source in the \emph{Spritz} round of the LDCs, and are listed in the third column of Table~\ref{tab:parameters-priors-truths}.
The system under consideration is also one of the 15 MBHB systems present in the previous \emph{Sangria} round.
This is a loud but short-lived MBHB signal, with an ideal SNR (as computed via Eq.~\eqref{eq:snr}, using the PSD defined in Appendix~\ref{sec:noise-psd}) of over 4000, mostly accumulated around the merger.

Our algorithms have been tested with both noise-free and noisy data $\vec{x}_{\mathrm{o}}$.
Because it was generated with our simulation pipeline, we know that the test data is generated using the same LISA orbits, TDI version and parameter conventions as the data that is used to train our models.
In order to apply this method to the LDCs, every part of the pipeline needs to be carefully consolidated -- the test data needs to be transformed to comply with our data format (i.e. cropping, downsampling and whitening it; performing dimensionality reduction, etc.) and our simulation pipeline needs to be reconfigured to better suit the data, including being able to generate non-stationary noise.
For each scenario, we have performed three different analyses:

\begin{description}[leftmargin=*]
\item[Eryn] This first run serves as a control mechanism to ensure the posteriors obtained are reasonable, using the likelihood-based methods described in Sec.~\ref{sec:data-analysis-basics}.
        It consists of a simple parallel-tempered MCMC search using the Eryn sampler on the likelihood function in Eq.~\eqref{eq:likelihood}.
        No dimensionality reduction is performed as part of the data pipeline\footnote{Actually, the signal has been whitened with the transformation of Eq.~\eqref{eq:whitening-transform}, but then Eq.~\eqref{eq:likelihood} has been modified to take this into account.}.

        We sample from 10 temperatures: the lowest one corresponds to $\beta = 1$, the target posterior, and the highest one to $\beta = 0$, the prior.
        The 8 remaining temperatures are initialized with a geometric spacing and are allowed to evolve during the sampling to ensure a 25\% swap probability between adjacent temperatures~\cite{Vousden:2016eeu}.
        Each temperature contains an ensemble of 64 MCMC chains being sampled in parallel.
        Eryn supports a variety of proposals for an MCMC jump, allowing for fine-tuning to the problem in question, but here we only use the affine-invariant stretch proposal \cite{Goodman:2010dyf,Foreman-Mackey:2012any}.

\item[SNL-PCA] The main method this work proposes makes use of the simulation pipeline described in Sec.~\ref{sec:sim-pipeline} to provide training data to the SNL algorithm presented in Sec.~\ref{sec:snl}.
        The code implementation of the algorithm relies on the \texttt{sbi} package~\cite{2020JOSS....5.2505T,tejero-canteroSbiToolkitSimulationbased2022}, which provides the backend and the building blocks required for the inference algorithm.
        In our experiments, we use a MAF consisting of 15 MADE transforms.
        Each MADE is a neural network containing 2 hidden layers with 128 neurons each.
        The algorithm has been left to train for 100 rounds, with $10000$ new simulations appended to the training dataset after each round.

        In order to make the inference more reliable regardless of the specific choice of parameters $\vecg{\theta}$, an embedding network has been used to transform $\vecg{\theta}$ before feeding it to the MAF.
        This consists of a shallow network, with 128 neurons in its hidden layer, that takes the parameter vector $\vecg{\theta}$ as input and outputs a transformed vector $\vecg{\theta}$ of the same length.
        The weights and biases of this hidden layer are trained at the same time as our MAF.

        Between SNL training rounds, new training data needs to be generated.
        For this, we use Eryn again to sample the posterior derived from the estimated likelihood obtained at the end of each round.
        We then run the simulator for the parameters $\vecg{\theta}$ obtained from this sampling procedure.
        We use the same 10-temperature scheme mentioned in the description of the Eryn analysis just above, but now each temperature runs 100 chains in parallel.
        The first round of SNL initializes these chains from the prior, but the state of the chains is stored so the chains in the next round can be initialized at the last position of each round.
        This procedure is expected to minimize the need of burn-in beyond the first round, but at the start of each round 1000 steps of burn-in are performed in any case so that the chains and temperatures can adapt to the new likelihood function.
        To ensure the samples are thoroughly uncorrelated, the posterior is sampled for 10 times the required number of steps, and the chains are thinned accordingly -- so that one in every 10 samples are used as training data for the next round.

\item[SNL-AE] Finally, we show results that use the modifications to the data generation pipeline suggested in Sec.~\ref{sec:modified-pipeline}.
        The configuration of the SNL algorithm is the same as in the SNL-PCA above, but the pipeline scales the data and feeds it through an autoencoder instead of using PCA.
        The autoencoders used have been pre-trained independently before doing any inference, as it has been described in Sec.~\ref{sec:autoencoders}.
        For the noise-free scenario, we use a standard autoencoder, while for the noisy case we use a denoising autoencoder.

        It is important to remark that, because the autoencoder architectures we have trained fail to reconstruct the merger, where most of the SNR is located, we do not expect the resulting posteriors to be as good as the other two methods.
        Nevertheless, we find that the results obtained are interesting enough that they deserve commentary.
\end{description}

The results for each of the two scenarios, noise-free and noisy, are discussed in the following subsections.

\subsection{\label{sec:results-noisefree}Noise-free case}

\begin{figure*}
    \includegraphics[]{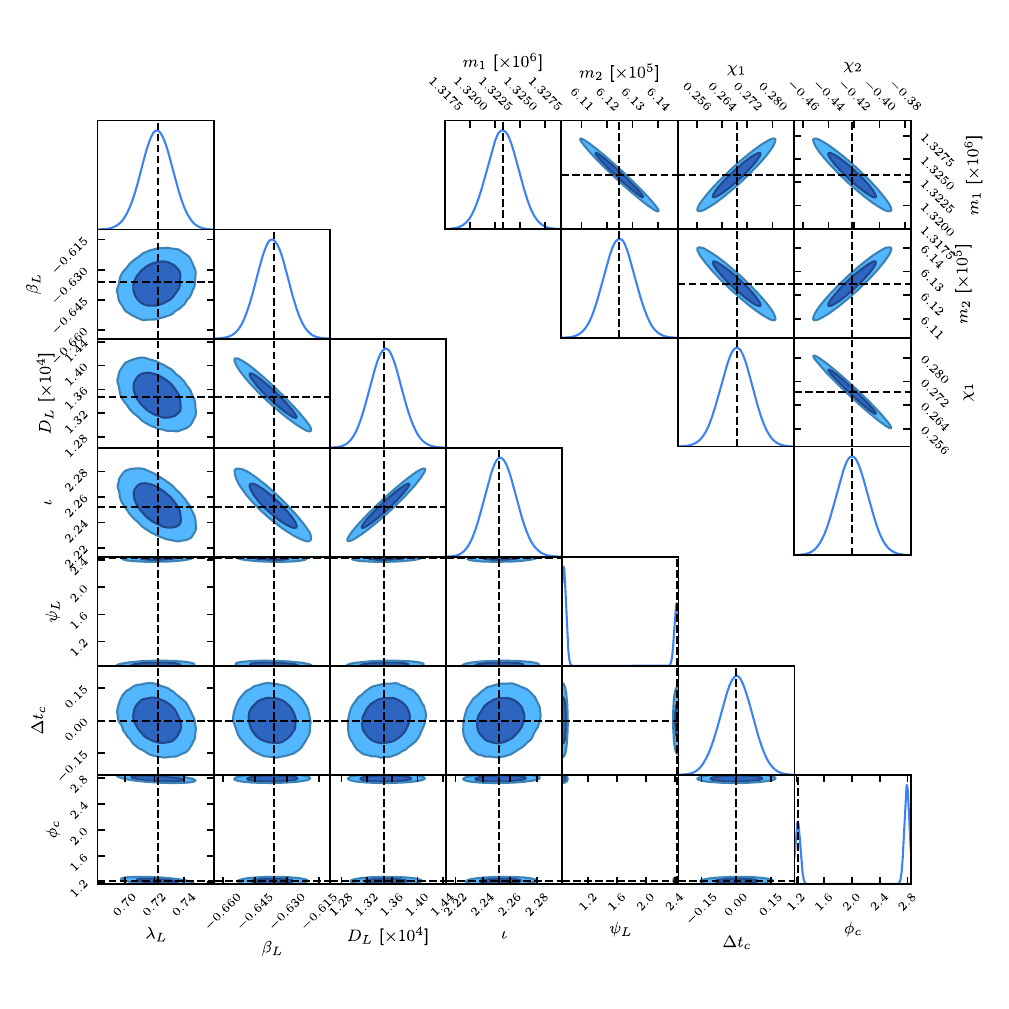}
    \caption[]{\label{fig:mcmc-results-noisefree-eryn} Eryn run, noise-free. The top right triangle represents the intrinsic parameters, while the bottom right contains the extrinsic ones. The~diagonal subplots in each triangle represent the 1D marginalized posterior distributions obtained for each parameter. The~off-diagonal contours show the 68\% and 95\% confidence intervals in the 2D marginalized posteriors for each combination of parameters. The~dashed black lines are the true parameter values of the injected signal.}
\end{figure*}

\begin{figure*}
    \centering
    \includegraphics[]{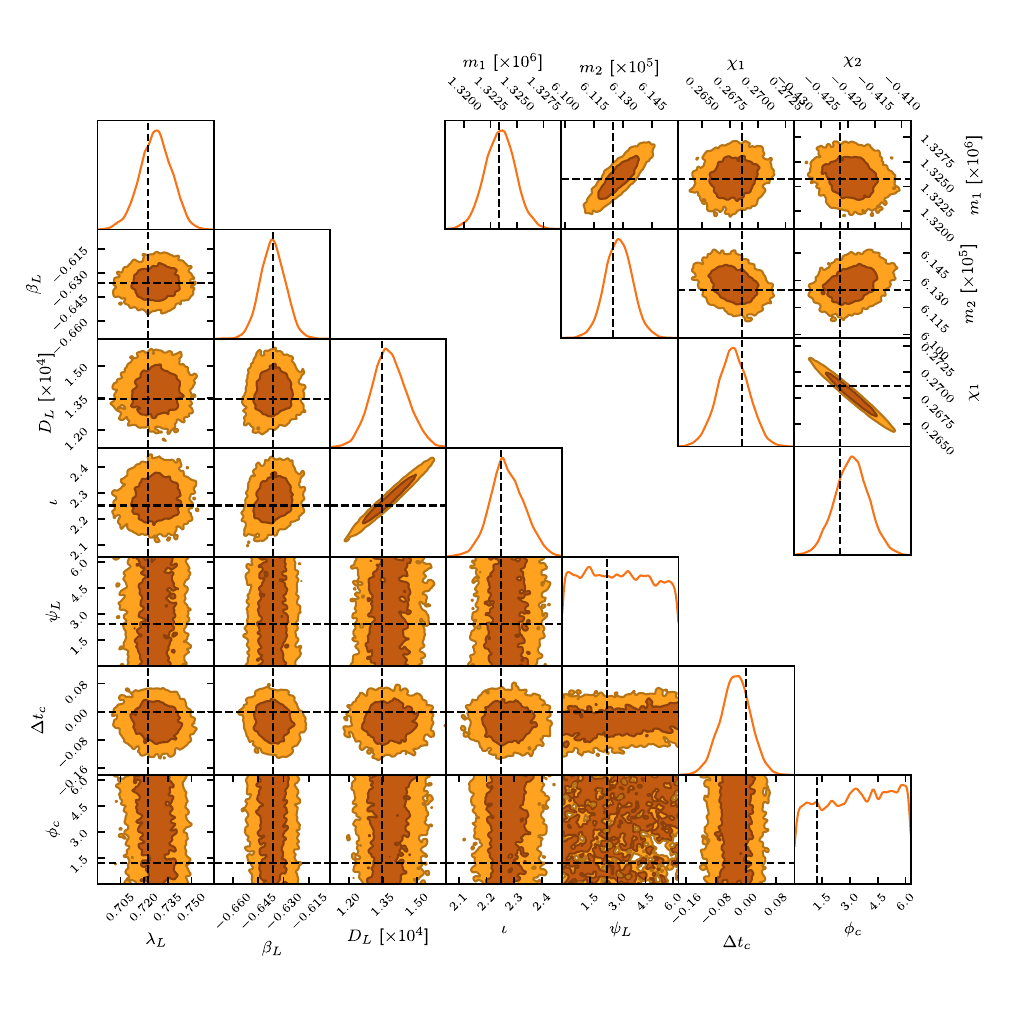}
    \caption{\label{fig:mcmc-results-noisefree-pca}SNL-PCA, noise-free. The top right triangle represents the intrinsic parameters, while the bottom right contains the extrinsic ones. The~diagonal subplots in each triangle represent the 1D marginalized posterior distributions obtained for each parameter. The~off-diagonal contours show the 68\% and 95\% confidence intervals in the 2D marginalized posteriors for each combination of parameters. The~dashed black lines are the true parameter values of the injected signal.}
\end{figure*}

\begin{figure*}
    \centering
    \begin{subfigure}{0.328\textwidth}
        \centering
        \includegraphics[width=\linewidth]{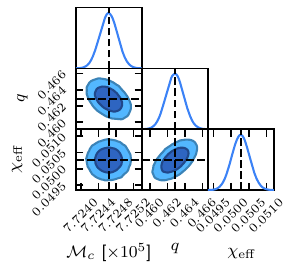}
        \caption{\label{fig:mcmc-results-noisefree-intrinsicextra-eryn}Eryn}
    \end{subfigure}
    \begin{subfigure}{0.328\textwidth}
        \centering
        \includegraphics[width=\linewidth]{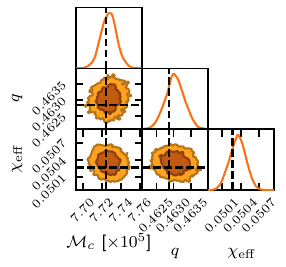}
        \caption{\label{fig:mcmc-results-noisefree-intrinsicextra-pca}SNL-PCA}
    \end{subfigure}
    \begin{subfigure}{0.328\textwidth}
        \centering
        \includegraphics[width=\linewidth]{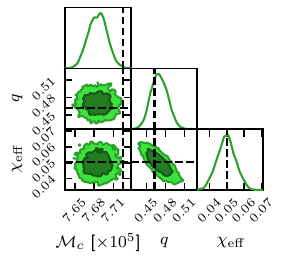}
        \caption{\label{fig:mcmc-results-noisefree-intrinsicextra-ae}SNL-AE}
    \end{subfigure}
    \caption{\label{fig:mcmc-results-noisefree-intrinsicextra} Noise-free posteriors for the chirp mass $\mathcal{M}_c$, the mass ratio $q$ and the spin combination $\chi_{\mathrm{eff}}$.}
\end{figure*}

We begin by showing the results in the absence of noise. Fig.~\ref{fig:mcmc-results-noisefree-eryn} shows the parameters recovered by Eryn and Fig.~\ref{fig:mcmc-results-noisefree-pca} shows the results obtained with SNL-PCA.
In both cases, for the intrinsic parameters --- as shown in the top right sections of each --- the recovered posteriors do have a maximum around the true parameter values, so they are correctly recovered by both algorithms.
The 1D projections for each intrinsic parameter are ultimately of similar width, with SNL-PCA being only slightly narrower than Eryn. 
The most notable difference between both posteriors is in the $m_1$--$m_2$ 2D posteriors, which have different shapes.
In order to diagnose this, we show in Fig.~\ref{fig:mcmc-results-noisefree-intrinsicextra} the posteriors after reparameterizing to the chirp mass $\mathcal{M}_c$, the mass ratio $q = m_2/m_1$, and the effective spin $\chi_{\mathrm{eff}}$ (see Eqs.~\eqref{eq:chirp-mass} and \eqref{eq:chi-effective-definition}).
Comparing Figs.~\ref{fig:mcmc-results-noisefree-intrinsicextra-eryn}~and~\ref{fig:mcmc-results-noisefree-intrinsicextra-pca}, we can see that SNL-PCA has not been able to recover the chirp mass $\mathcal{M}_c$ with the same accuracy as Eryn, but in turn, it constrained the mass ratio $q$ remarkably better.
Thus, the elongated shape of the $m_1$-$m_2$ posterior in Fig.~\ref{fig:mcmc-results-noisefree-eryn} approximately follows a line of constant $\mathcal{M}_c$, while the corresponding panel in Fig.~\ref{fig:mcmc-results-noisefree-pca} is more distributed along the direction of constant $q$.
The effective spin parameter $\chi_{\mathrm{eff}}$, as defined in Eq.~\eqref{eq:chi-effective-definition}, depends only on $q$ and the individual spins $\chi_1$ and $\chi_2$ --- the fact that SNL-PCA is able to constrain these three parameters better than Eryn explains the narrower effective spin posteriors.

Qualitatively similar results are also obtained for the extrinsic parameters, as shown in the lower left triangles of Figs.~\ref{fig:mcmc-results-noisefree-eryn} (Eryn) and~\ref{fig:mcmc-results-noisefree-pca} (SNL-PCA).
There are some differences between them: while SNL-PCA does not recover the inclination or luminosity distance as well as Eryn, SNL-PCA is able to place a much tighter constraint on the longitude angle.

It should be noted that, during training, SNL-PCA was not able to resolve the longitude $\lambda_L$ well for several rounds, with the 1D posteriors looking similar to the priors.
This situation continued until, after having narrowed down most of the other parameters, the neural network training hit a sudden `breakthrough'.
In the particular run displayed in Fig.~\ref{fig:mcmc-results-noisefree-pca}, this did not happen until training round 33 -- until then, SNL-PCA seemed unable to learn about the longitude parameter.
From then on, the $\lambda_L$ marginal posterior estimation rapidly improved until converging to the one shown in Fig.~\ref{fig:mcmc-results-noisefree-pca}, which is narrower than the one recovered by Eryn.

The most obvious difference between the two posteriors is that SNL-PCA was not able to determine the polarization angle and the phase at coalescence. Meanwhile, Eryn finds a bimodal structure in these two quantities.
It is unclear to us whether the small loss of information caused by the dimensionality reduction employed in SNL could play a role in this discrepancy, or if simply letting the algorithm train for more than 100 rounds would resolve the issue, in a similar manner as with the longitude.

\begin{figure*}
    \includegraphics[]{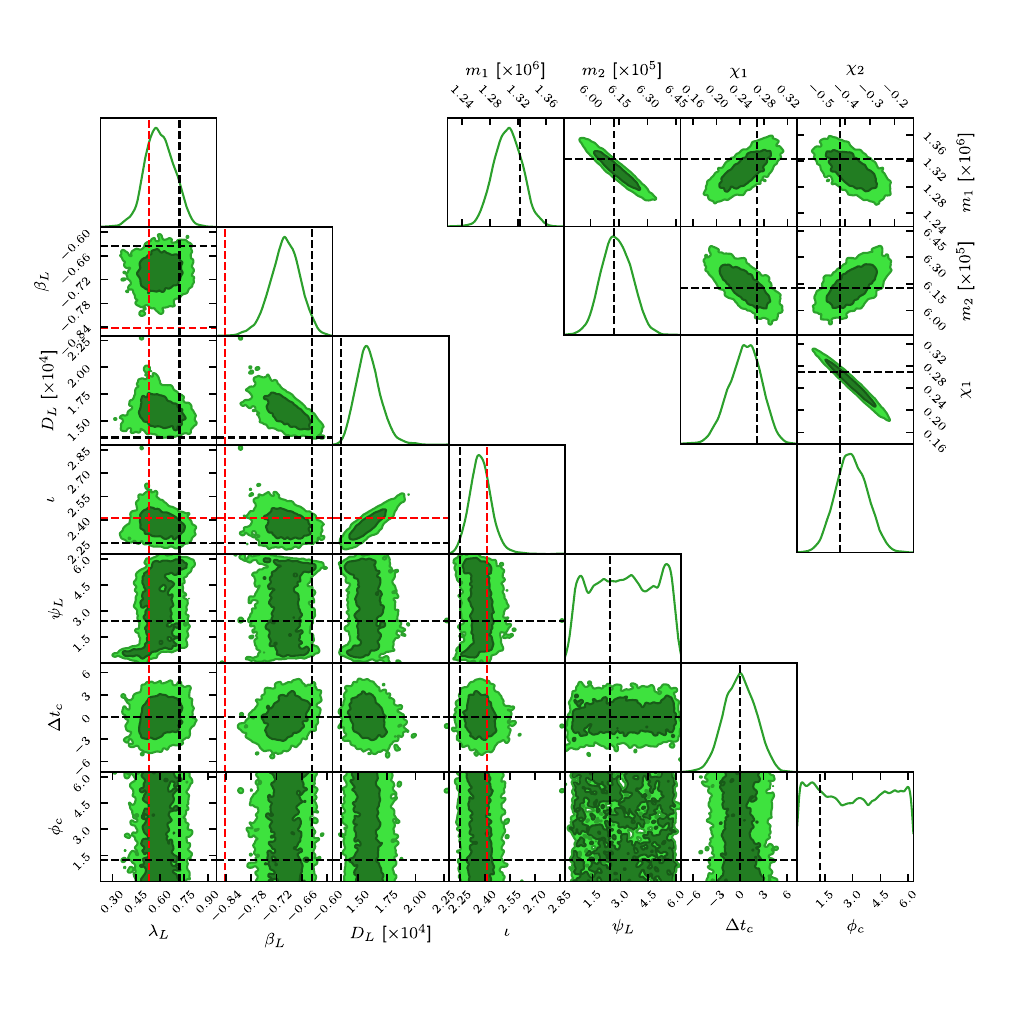}
    \caption{\label{fig:mcmc-results-noisefree-ae}SNL-AE, noise-free. The top right triangle represents the intrinsic parameters, while the bottom right contains the extrinsic ones. The~diagonal subplots in each triangle represent the 1D marginalized posterior distributions obtained for each parameter. The~off-diagonal contours show the 68\% and 95\% confidence intervals in the 2D marginalized posteriors for each combination of parameters. The~dashed black lines are the true parameter values of the injected signal. The additional dashed red lines are the biased values of $\lambda_L$, $\beta_L$ and $\iota$, as predicted by Ref.~\cite{Marsat:2020rtl}.}
\end{figure*}

As for SNL-AE, Fig.~\ref{fig:mcmc-results-noisefree-ae} shows the resulting posterior.
We can see that our algorithm is able to successfully extract the intrinsic parameters of the injected MBHB source. The results in these parameters show posteriors noticeably broader than Eryn (Fig.~\ref{fig:mcmc-results-noisefree-eryn}), but otherwise they are qualitatively identical.
The broadening of the posterior for the intrinsic parameters translates into much wider posteriors for the intrinsic parameter combinations in Fig.~\ref{fig:mcmc-results-noisefree-intrinsicextra-ae}.

Regarding the extrinsic parameters, SNL-AE has also been able to recover them correctly, with a similar qualitative look to SNL-PCA (Fig.~\ref{fig:mcmc-results-noisefree-pca}) but with much broader posteriors.
There seems to be a bias in the sky location parameters $(\lambda_L, \beta_L)$ and the inclination $\iota$ with respect to their true values.
Ref.~\cite{Marsat:2020rtl} shows that, for short-lived low-frequency signals, the LISA response shows an apparent bias in these parameters.
This bias, in turn, induces biases in the other extrinsic parameters, caused by the correlations between parameters\footnote{For example, the luminosity distance $D_L$ is correlated with $\iota$, and a bias in the latter introduces a bias in the former.}.
In principle, the GW signal we are analyzing should not be in the regime where these biases appear.
However, as discussed in Sec.~\ref{sec:autoencoders} and illustrated in Figs.~\ref{fig:reconstruction-ae}~and~\ref{fig:reconstrunction-dae}, the autoencoders we train are zero-ing out the merger, unintentionally putting the signal in the low-frequency ($f \ll f_L = 1/L = \qty{0.12}{\hertz}$) regime.
The dashed red lines in Fig.~\ref{fig:mcmc-results-noisefree-ae} show the position of the biased true values as predicted by Ref.~\cite{Marsat:2020rtl} for the sky location and $\iota$.
For all three parameters, the maximum in the marginalized posteriors lies somewhere in between the true value and the predicted bias.
While the parameters obtained are biased, the predicted bias was somewhat suppressed, presumably because the signal is still not quite in the short-lived low-frequency regime.

It should be noted that, during the training process of SNL-AE, a secondary mode in the sky location and inclination appeared in the posteriors.
This secondary mode is compatible with the transformation
\begin{equation}
    \lambda^{}_L \rightarrow \lambda^{}_L - \pi\,,\quad
    \beta^{}_L \rightarrow -\beta^{}_L\,, \quad
    \iota \rightarrow \pi - \iota\,,
\end{equation}
which is the \emph{antipodal} mode described in Ref.~\cite{Marsat:2020rtl}.
There, it is shown that this particular mode appears in the form of a degeneracy when the LISA response function is approximated in the low-frequency regime we just have described.
The simulator in our algorithm uses the full LISA response, so the presence of this mode may seem surprising.
But it can also be explained by the autoencoder failure to reconstruct the merger.
That is, the low-dimensional summaries used in the inference algorithm do not carry the high-frequency information necessary to reconstruct the merger.
This effectively puts our data near the regime of the low-frequency approximation.
When using the leading-order low-frequency response, the two modes should have the same amplitude, but higher-order corrections decrease the amplitude of the secondary mode.
This secondary mode is not visible in Fig.~\ref{fig:mcmc-results-noisefree-ae}.
This is because, by the end of the training, only 13 of the 10000 posterior samples used in the elaboration of the figures could be considered to belong to it.
The secondary mode is still being explored in the higher-temperature posterior chains, however.

SNL-AE managed to recover the merger time within 10 seconds.
This is not as good as the Eryn or SNL-PCA, but considering that the dimensionality reduction is effectively inserting a data gap during the merger itself, the fact that our algorithm was still able to recover $\Delta t_c$ with an accuracy within a time-domain bin can be considered a remarkable success and an indicator of the robustness of SNL to artifacts, as long as the training data also contains them.
Indeed, it is also important to mention that SNL-PCA was able to obtain a precision in $\Delta t_c$ comparable to Eryn, considering that the data being fed to the algorithm is essentially heavily compressed time-domain data with a low sample rate.
On the other hand, SNL-AE, like SNL-PCA, is also unable to satisfactorily determine $\psi$ and $\phi_c$.

Our experiments determine empirically that the number of simulations that are run between each round of inference can play an important role in the speed of training.
This hyperparameter effectively dictates the relative amount of time that should be spent on sampling from the posterior and running simulations compared to neural network training.
For a fixed amount of simulations, spreading them into several smaller rounds seems to lead to higher quality posteriors than doing fewer rounds with more simulations.
This also seems to net decent posteriors in a relatively short time, but the rate of improvement stagnates earlier.
Conversely, if the main goal is to obtain high quality posteriors at the end of training, performing fewer training rounds with more simulations between each seems to yield better results.
This can be explained by the fact that, with smaller rounds, fewer simulations are `spent' in the non-informative prior early on.
However, once the likelihood estimate has started to converge, adding fewer simulations to the training data in each round means a slower rate of progress per round.
In later stages of the training, when the training dataset is already large and each round of training becomes computationally expensive, improvements become very marginal and many rounds are needed.
With larger rounds, more resources are spent early on simulating waveforms in parts of parameter space that are not as relevant, but the likelihood estimate will be able to be trained further before the rate of progress begins to stagnate.

\subsection{Noisy case}

\begin{figure*}
    \includegraphics[]{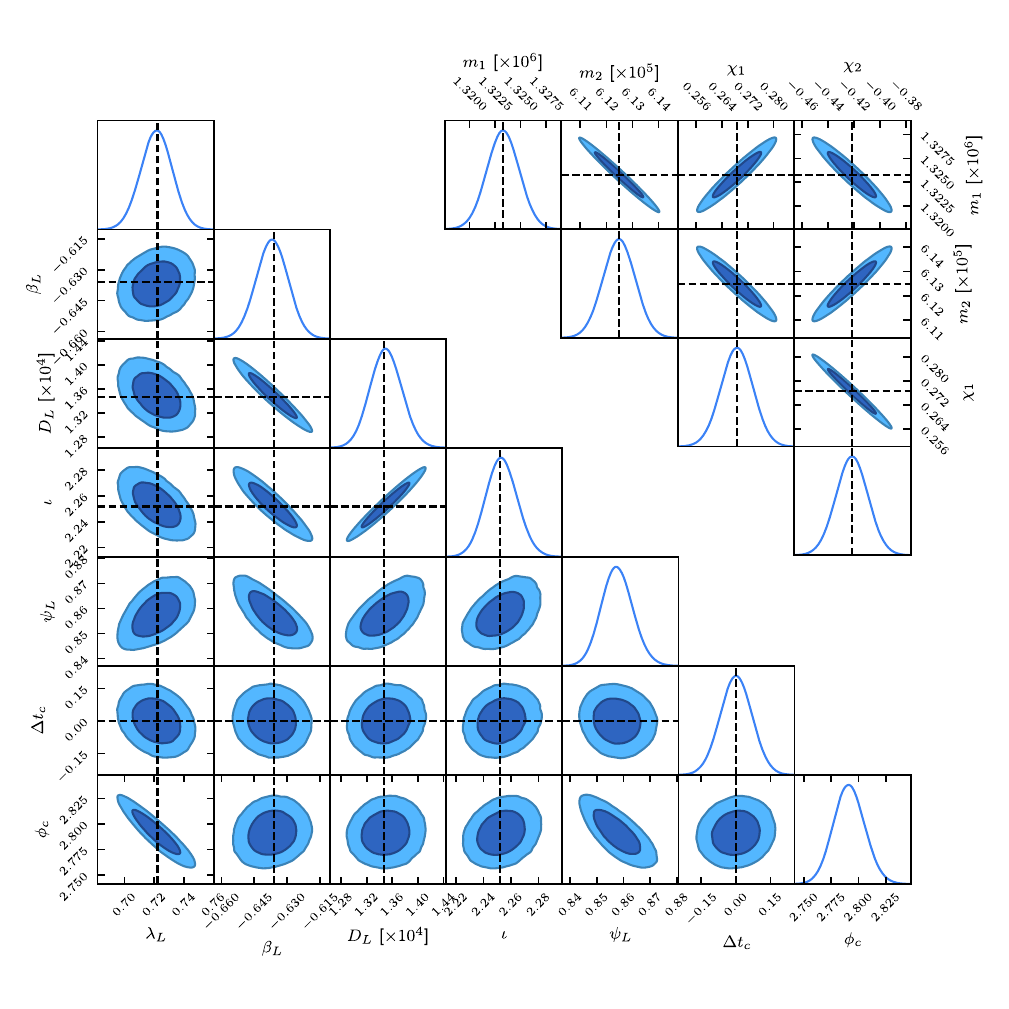}
    \caption{\label{fig:mcmc-results-noisy-eryn}Eryn, noisy. The top right triangle represents the intrinsic parameters, while the bottom right contains the extrinsic ones. The~diagonal subplots in each triangle represent the 1D marginalized posterior distributions obtained for each parameter. The~off-diagonal contours show the 68\% and 95\% confidence intervals in the 2D marginalized posteriors for each combination of parameters. The~dashed black lines are the true parameter values of the injected signal.}
\end{figure*}

\begin{figure*}
    \includegraphics[]{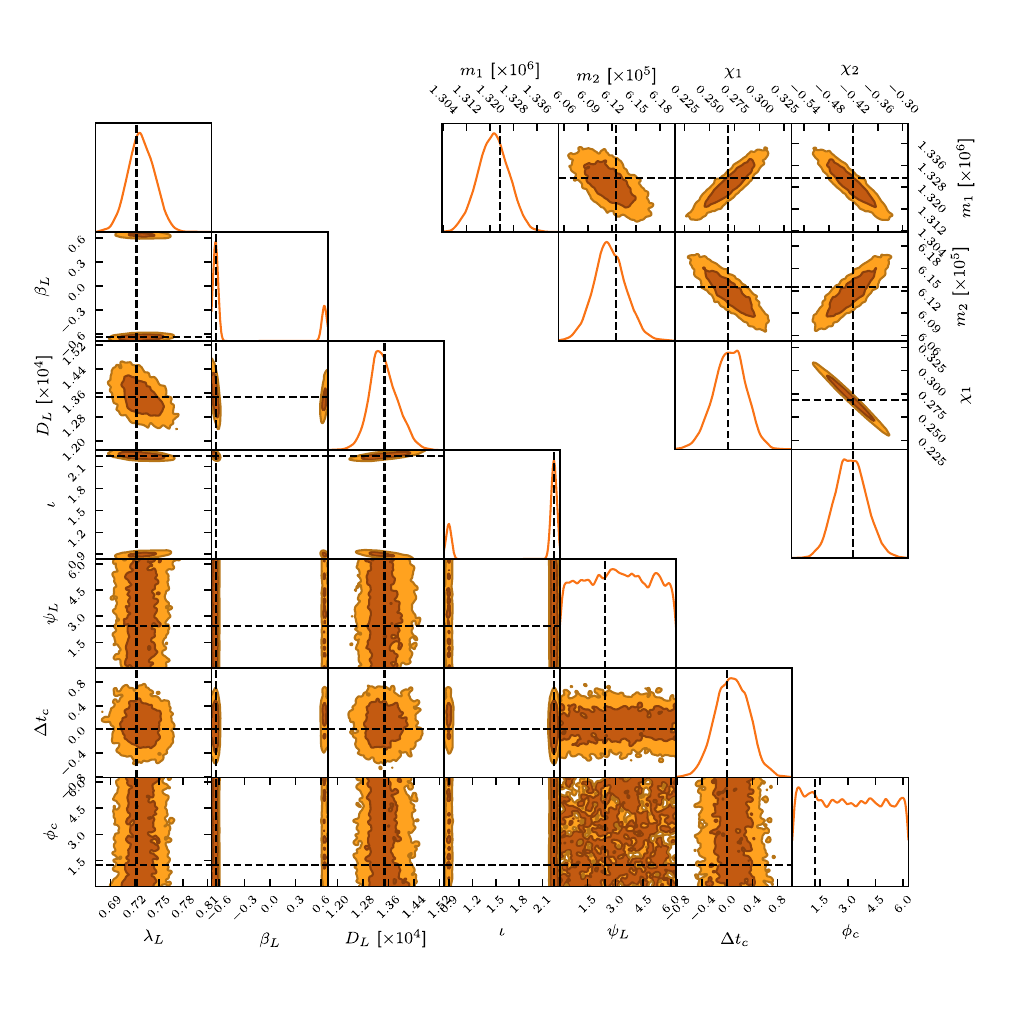}
    \caption{\label{fig:mcmc-results-noisy-pca}SNL-PCA, noisy. The top right triangle represents the intrinsic parameters, while the bottom right contains the extrinsic ones. The~diagonal subplots in each triangle represent the 1D marginalized posterior distributions obtained for each parameter. The~off-diagonal contours show the 68\% and 95\% confidence intervals in the 2D marginalized posteriors for each combination of parameters. The~dashed black lines are the true parameter values of the injected signal.}
\end{figure*}

\begin{figure*}
    \centering
    \begin{subfigure}{0.328\textwidth}
        \centering
        \includegraphics[width=\linewidth]{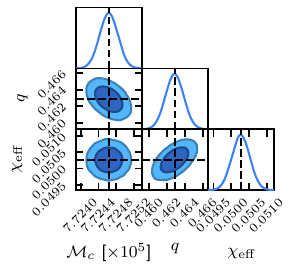}
        \caption{\label{fig:mcmc-results-noisy-intrinsicextra-eryn}Eryn}
    \end{subfigure}
    \begin{subfigure}{0.328\textwidth}
        \centering
        \includegraphics[width=\linewidth]{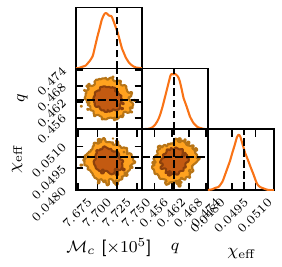}
        \caption{\label{fig:mcmc-results-noisy-intrinsicextra-pca}SNL-PCA}
    \end{subfigure}
    \begin{subfigure}{0.328\textwidth}
        \centering
        \includegraphics[width=\linewidth]{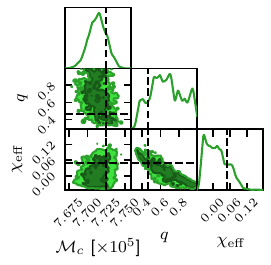}
        \caption{\label{fig:mcmc-results-noisy-intrinsicextra-ae}SNL-AE}
    \end{subfigure}
    \caption{\label{fig:mcmc-results-noisy-intrinsicextra} Noisy posteriors for the chirp mass $\mathcal{M}_c$, the mass ratio $q$ and the spin combination $\chi_{\mathrm{eff}}$.}
\end{figure*}

The presentation of the results with noise follows the same line as in the noise-free case.
Fig.~\ref{fig:mcmc-results-noisy-eryn} shows the parameters recovered by Eryn, while Fig.~\ref{fig:mcmc-results-noisy-pca} shows the SNL-PCA equivalent.
Comparing them, we can see qualitatively similar results for the intrinsic parameters, where the posteriors have similar shapes and are centered around the true values, but the SNL-PCA posteriors are slightly broader due to lower fidelity in recovering the chirp mass.
Fig.~\ref{fig:mcmc-results-noisy-intrinsicextra} shows the intrinsic posteriors of the chirp mass $\mathcal{M}_c$, the mass ratio $q$ and the effective spin $\chi_{\mathrm{eff}}$.
Here, we see that the SNL-PCA posteriors of Fig.~\ref{fig:mcmc-results-noisy-intrinsicextra-pca} remain compatible with the truth --- although they are slightly broader than Eryn's (Fig.~\ref{fig:mcmc-results-noisy-intrinsicextra-eryn}), and the correlation between parameter pairs is lost.

As for the extrinsic parameters, for those for which the behavior is qualitatively similar in both algorithms, the SNL-PCA posteriors are still slightly broader than the Eryn ones.
However, there are a few notable differences, both between each other and with the noise-free case.
First, Eryn seems to not have caught the secondary modes in $\psi$ and $\phi_c$, focusing only on the incorrect one, although the algorithm is able to correctly determine the remaining 9 parameters.
On the other hand, as it happens in the noise-free case, SNL-PCA is unable to determine these two parameters.
More interestingly, the SNL-PCA posteriors feature two modes in the extrinsic parameters. They are related to each other by the transformation
\begin{equation}
    \lambda^{}_L \rightarrow \lambda^{}_L \,, \quad
    \beta^{}_L \rightarrow - \beta^{}_L \,, \quad
    \iota \rightarrow \pi - \iota \,.
\end{equation}
This secondary mode is what Ref.~\cite{Marsat:2020rtl} refers to as the \emph{reflected} sky position mode.
This mode arises from a degeneracy in the LISA response function under the approximation of a static configuration of the spacecrafts.
Of course, this work is using Keplerian orbits, but for a signal that is short-lived enough, LISA does not have time to move much and this regime applies.
The observation durations we are considering are long enough that LISA can move a few degrees and break this degeneracy, but the SNR of this particular MBHB GW source is so focused on the merger that it is effectively in this low-duration regime.
This mode only seems to appear when the data is noisy because the noise effectively masks the inspiral part of the waveform, shortening the effective duration of the signal.
In early rounds of SNL training, the two modes first appear with similar amplitude, but as rounds pass the secondary mode begins getting suppressed.
There is the possibility that with more rounds of training, the reflected mode may disappear altogether.
While this secondary mode does not appear in Eryn (Fig.~\ref{fig:mcmc-results-noisy-eryn}), we should note that it is sampled from briefly during the burn-in MCMC iterations, and the higher-temperature chains do explore this mode.

\begin{figure*}
    \includegraphics[]{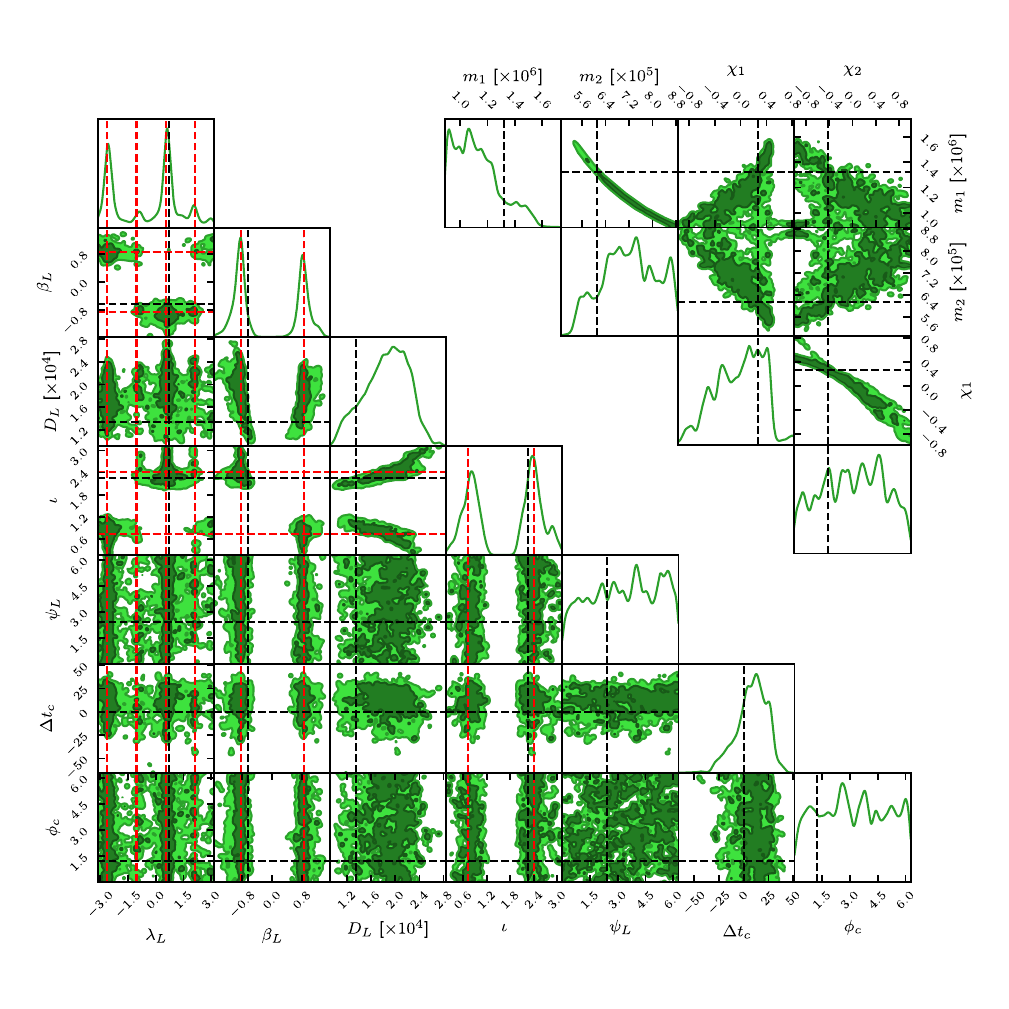}
    \caption{\label{fig:mcmc-results-noisy-ae}SNL-AE, noisy. The top right triangle represents the intrinsic parameters, while the bottom right contains the extrinsic ones. The~diagonal subplots in each triangle represent the 1D marginalized posterior distributions obtained for each parameter. The~off-diagonal contours show the 68\% and 95\% confidence intervals in the 2D marginalized posteriors for each combination of parameters. The~dashed black lines are the true parameter values of the injected signal.
        The additional dashed red lines are the biased values of $\lambda_L$, $\beta_L$ and $\iota$, as predicted by Ref.~\cite{Marsat:2020rtl}.}
\end{figure*}

Finally, Fig.~\ref{fig:mcmc-results-noisy-ae} shows the posterior obtained for SNL-AE.
Extraction of the intrinsic parameters from the posteriors becomes a lot more difficult for SNL-AE.
For the masses, while the 1D marginalized posteriors may be off from their true value, in the $m_1$-$m_2$ projection we can see that the estimated masses are very well localized on a curve of constant chirp mass: the chirp mass is well recovered, in contrast with  the mass ratio.
Indeed, Fig.~\ref{fig:mcmc-results-noisy-intrinsicextra-ae} shows that the one-dimensional posterior for the chirp mass $\mathcal{M}_c$ is of a similar width and shape to the one recovered by SNL-PCA (Fig.~\ref{fig:mcmc-results-noisy-intrinsicextra-pca}), while the mass ratio $q$ remains badly constrained.
The individual spins, $\chi_1$ and $\chi_2$, are also poorly recovered --- as can be seen in Fig.~\ref{fig:mcmc-results-noisy-ae}, while the mode of $\chi_1$ is somewhat near the true value, the marginal posteriors for both parameters occupy nearly the entire range of possible values.
However, the projection in the $\chi_1$-$\chi_2$ plane shows that the spins are rather well-determined in a combination of $\chi_1$ and $\chi_2$ --- just like the chirp mass was compared to the individual mass posteriors.
The $\chi_{\mathrm{eff}}$ posterior, shown in Fig.~\ref{fig:mcmc-results-noisy-intrinsicextra-ae}, is significantly narrower than either of the individual spins, but the shape remains irregularly-shaped, biased from the true value, and much broader than both methods. 
It should be noted that the upper right corner of Fig.~\ref{fig:mcmc-results-noisy-ae} looks qualitatively similar to Fig.~8 of Ref.~\cite{Cornish:2020vtw}.
This figure shows the posterior for the intrinsic parameters for a pre-merger MBHB signal -- that is, including only the inspiral part of the waveform.
The fact that our results look similar to that scenario reinforces the idea that the degradation in our SNL-AE posteriors can be explained by the fact that our autoencoder is failing to reconstruct the merger, as explained in Sec.~\ref{sec:autoencoders} (and shown explicitly in Figs.~\ref{fig:reconstruction-ae}~and~\ref{fig:reconstrunction-dae}).

Regarding the extrinsic parameters, the lower left corner of Fig.~\ref{fig:mcmc-results-noisy-ae} shows that, for most parameters, the qualitative behaviour is the same as in the noise-free case (Fig.~\ref{fig:mcmc-results-noisefree-ae}), albeit with broader posteriors.
For the sky location, instead of one or two modes, we see a more complex structure with at least 6 distinguishable modes.
These modes are all compatible with the 8 modes that \cite{Marsat:2020rtl} predicts for coalescing binaries in the limit of short-lived low-frequency signals.
Indeed, the modes are located precisely where Ref.~\cite{Marsat:2020rtl} predicts.
The two dominant modes are the true and antipodal modes mentioned in Sec.~\ref{sec:results-noisefree}, with the remaining four being suppressed.
We hypothesize that the modes coming from reflections $\beta_L \rightarrow -\beta_L$ of the two dominant ones are even more suppressed, to the point that the MCMC run used in elaborating Fig.~\ref{fig:mcmc-results-noisy-ae} captured just a few samples near the reflection of the true mode, and none in the reflection of the antipodal one.

\section{Conclusions and Discussion\label{sec:discussion}}
In this work, we have investigated for the first time how to use NLE methods for the detection and parameter estimation of MBHB signals in synthetic LISA data.
Despite the complex structure of the posterior distributions, with multi-modalities and degeneracies in various parameters, we have shown that MAF-based methods are able to successfully recover them.
The implementation of SNL presented in this paper contains various tunable components that are individually complex enough to deserve dedicated investigations to optimally integrate them in the global scheme of our method. The list of ingredients includes: (i) Choice of the MBHB waveform model (we can add higher modes and/or precessional effects); (ii) data representation (frequency versus time-domain versus time-frequency representations); (iii) the choice of posterior sampler (MCMC variants or other sampling algorithms) and its corresponding hyperparameters; (iv) the choice of data compression technique (PCA, autoencoders, etc.); and (v) the choice of neural network architecture for the normalizing flows and autoencoders, if the latter are being used. All this means that there is significant room for improvement of the techniques presented in this paper, which is necessary in order to make them competitive with respect to more established methods like MCMC.

Nevertheless, the adaptive part of our algorithm means we are already more efficient than other methods in terms of the number of GW waveform model evaluations.
As an advantage over other methods, all assumptions on the noise and the physics included in the waveform model are implicitly given by the simulation pipeline used to generate the training data.
This means that, including waveforms with more physics, or nonstationary noise requires no modification on the inference algorithm --- provided that the information gained from this added complexity survives the dimensionality reduction step.
In fact, since data is being dynamically generated while targeting an observation, it would be possible to begin training with simpler simulations, then fine-tune with more complex ones.

A key consideration for these methods to be feasible is dimensionality reduction of the data. Special care is required to minimize the loss of information in this process. For the particular scenario we deal with in this paper, PCA to $128$ components (using the procedure described in Sec.~\ref{sec:sim-pipeline}) yields good results -- though they are not fully identical to the likelihood-based results.
In anticipation for the complications that are going to appear in more realistic scenarios, including glitches, gaps and non-stationarities in the data, in Sec.~\ref{sec:modified-pipeline} we have proposed some modifications and introduced autoencoders as an alternative dimensionality reduction procedure.
Autoencoders are a type of neural network that is very flexible. In this paper we have only investigated simple configurations, so that the associated architectures that we have trained on scaled time-series data have not been able to resolve the merger part of the signal. Nevertheless, the results we have obtained look quite promising.

Simulation-based inference methods give us freedom in terms of the representation of the signal before data compression. In this work, we have only used signals in the time domain -- unscaled for the SNL-PCA method, and scaled for autoencoders\footnote{We have also trained autoencoders with unscaled data, but in that case our architectures are only able to reconstruct the merger itself. The resulting posteriors are better than with scaled data, but not as good as SNL-PCA, so we do not consider them to be interesting enough.}.
There is still a lot of room for more experimentation, in particular with other representations of the data. For instance, it may be more convenient to use frequency-domain data, or a combination of both, or two-dimensional representations such as the ones provided by wavelet methods.
These alternative representations may appear as more suitable for different dimensionality reduction algorithms. In this sense, the modular character of the simulation pipeline we have developed in this work can serve as a basis to experiment with these different representations.

It is important to remark that the particular arquitecture for the autoencoder used in the SNL-AE method (shown in Appendix~\ref{sec:ae-architecture}) has been obtained by a process of trial and error.
In principle, we have complete freedom in the definition of the autoencoder, beyond the one-dimensional convolutional neural network for which we have shown results in this work.
Indeed, we have also experimented with fully-connected autoencoders, but the large number of network parameters they involve have caused overfitting to the training data.
Nevertheless, it seems very likely that, with more experimentation and with the help of automated hyperparameter searches, we can find configurations that yield better results and in a more efficient way.
On the other hand, there is also the possibility of experimenting with other dimensionality reduction algorithms beyond PCA and autoencoders.

Regarding the choice of sampler, just as in the case of likelihood-based methods, SNL requires an algorithm that is resilient when multimodalities appear.
The original implementation of SNL~\cite{Papamakarios:2018zoy} recommends using slice sampling \cite{2000physics...9028N}, as it fits the criteria and little tuning is necessary.
However, multidimensional slice sampling is Gibbs-like in the sense that, at each iteration, a new value for each individual parameter is proposed conditioned on the current state of all the others.
In our problem, some of the parameters are correlated or even degenerate, which restricts significantly the size of the jumps that can be made and impedes the efficient navigation of the parameter space.
This could, in principle, be mitigated by redefining the parameters to be sampled in order to circumvent the correlations, but this lies outside of the scope of this article and will the subject of future investigations.
Instead, we find that reusing Eryn's parallel-tempered MCMC to sample the posterior between rounds is sufficient.

In terms of computational performance\footnote{The wall times reported here were measured with a 12-core AMD Ryzen 5900X CPU.}, with our choice of hyperparameters, our synthetic likelihood is evaluated in $\qty{3}{\milli\second}$.
This time saving is not much as compared to the $\qty{4}{\milli\second}$ that it takes to evaluate the full likelihood.
However, since our trained likelihood estimate does not require any simulations, it can be vectorized to a higher degree so that more evaluations can be performed in parallel: the MCMC runs done between SNL rounds took approximately 100 seconds to produce 10000 effective samples -- with 2 million total likelihood calls, accounting for the burn-in and thinning.
Each of our SNL runs have been restricted to a total of $\num{d6}$ waveform calls across all rounds, while the likelihood-based runs we compared against did $\num{6.4d7}$.
Thus, we are able to achieve qualitatively comparable results with fewer than 2\% of the simulations.
In the future, as the realism -- and cost -- of our simulations increases, by introducing higher-order waveform corrections, glitches and gaps, we expect the difference in efficiency to widen in our favor.

It must also be noted that the evaluation time of our synthetic likelihood strongly depends on the complexity of the underlying MAF.
We have some control over it, in the form of the number of MADEs that form it and the size and number of hidden layers within each MADE.
Indeed, experiments have shown us that reducing the number of MADEs from 15 to 5 improves the computational performance by a factor of 3 with no noticeable loss of posterior quality, showing that there is room for optimization.

The other way to modify the complexity of a MAF is by varying the number of inputs and outputs, which is tied to the dimensionality of the compressed data.
In this work, we have settled for 128 dimensions, but further improvements to dimensionality reduction would dramatically drive the cost down.
After all, there exists a representation of the full TDI data in as few as 11 numbers -- the parameters $\vecg{\theta}$ themselves.

In our experiments, we observe that SNL trains very quickly at first but slows down as rounds pass and the size of the training dataset grows.
Training a likelihood estimate for 100 rounds of 10000 simulations took approximately 80 hours.
This is rather slow -- much more so than the 6 hours it took to run Eryn -- but all the potential performance improvements we have just suggested would compound into lowering this cost.
Depending on the scenario, the improvement over the posteriors is not necessarily gradual -- it may seem like training has converged for a few rounds, only for the training to hit a breakthrough and suddenly improve drastically. In this sense, it is not clear  whether the parameters that SNL was unable to determine are going to be found with more SNL rounds.
The number of simulations appended to the training dataset in each round also plays an important role in the performance of the algorithm.
We have seen that training for many small rounds will result in slower training, but with fewer overall simulations needed.
Conversely, using only a few large rounds achieves equivalent results in a shorter time, but using a larger number of simulations.
Something that remains to be explored is the usage of a variable number of simulations per round -- for example, adding more simulations in later rounds, so the likelihood estimates can keep improving for more rounds.

Compared to other GW parameter estimation algorithms, both likelihood-based and simulation-based, the computational resources we have required are quite modest.
The dimensionality reduction step allows for the neural networks involved in the data analysis algorithm to be small and shallow enough that, for our relatively small-scale experiments, there are no significant performance gains in using GPUs.
Nevertheless, the use of GPUs for the training may improve performance by allowing us to increase the throughput of simulations shown to the inference networks with a corresponding increase in the training batch size -- with CPUs, the performance benefit of higher batch sizes is limited up to the number of available CPU cores.
GPUs are more directly beneficial in the training of the autoencoders, because they are much larger networks with more trainable parameters.
We have used a consumer-grade NVIDIA RTX 3080 GPU to train them.
The main limiting factor there has been the amount of GPU memory, which restricted the size of the model and the training dataset.

Although SNL does not have the capability to perform amortized inference, unlike NPE, the underlying neural networks for inference are much smaller and simpler.
Therefore, our computational requirements are significantly smaller -- in the required number of simulations, training time, and hardware -- than in methods like DINGO.
The downside is that we need to retrain our model for every new source.
Then, the scenario for which SNL is best suited is for systems where simulations have a high cost, and likelihoods and posteriors can be complicated, but we expect there to be relatively few independent sources.
In the case of LISA, the prime candidate sources are MBHBs: in the presence of higher modes, glitches, data gaps and non-stationary noise, the simulation cost is expected to be significantly higher.
The higher modes -- when resolvable -- eliminate many of the degeneracies in the posterior~\cite{Marsat:2020rtl} and simplify its overall structure, but the noise and data artifacts may reverse that effect to some degree~\cite{Dey:2021dem}.
Given that it is expected to have of the order of tens of MBHB detections per year, performing SNL for each of them as they arrive is feasible.
In a scenario where the rate of EMRI detections is comparable~\cite{Babak:2017tow}, this may also be a good method.
Finally, despite the performance and capabilities of the techniques and pipeline presented in this work, improvements are necessary in order for this method to provide performances comparable to methods as MCMC or NPE-based methods. This motivates to explore in future investigations the different avenues discussed above in which the method can be improved and its performance can be enhanced.

\begin{acknowledgments}
    We would like to thank the members of LDC working group of the LISA Consortium for fruitful discussions. We also thank Sascha Husa for feedback in the later stages of the elaboration of this article.
    IMV and CFS  are supported by contracts PID2019-106515GB-I00 and PID2022-137674NB-I00 from MCIN/AEI/10.13039/501100011033 (Spanish Ministry of Science and Innovation) and 2017-SGR-1469 and 2021-SGR-01529 (AGAUR, Generalitat de Catalunya).
    IMV has been supported by FPI contract PRE2018-083616 funded by MCIN/AEI/10.13039/501100011033 (Spanish Ministry of Science and Innovation) and “ESF Investing in your future”, and by NextGenerationEU/PRTR (European Union).
    This work has also been partially supported by the program \textit{Unidad de Excelencia Mar\'{\i}a de Maeztu} CEX2020-001058-M (Spanish Ministry of Science and Innovation).

    Our code has made use of \texttt{sbi} \cite{2020JOSS....5.2505T,tejero-canteroSbiToolkitSimulationbased2022} and \texttt{Eryn} \cite{Karnesis:2023ras,katzMikekatz04ErynFirst2023} for the SNL building blocks and MCMC implementations; \texttt{scikit-learn} \cite{Pedregosa:2011ork} for its implementation of PCA; \texttt{lisabeta} (see~\cite{Marsat:2018oam,Marsat:2020rtl})
    for the fast waveform implementation used in the simulation pipeline; \texttt{LDC} tools~\cite{LDCSoftware} for the implementation of the noise PSD described in Appendix~\ref{sec:noise-psd}; \texttt{Matplotlib} \cite{Hunter:2007ouj,teamMatplotlibVisualizationPython2024} and \texttt{ChainConsumer} \cite{2016JOSS....1...45H} for the elaboration of figures and visualizations; and \texttt{PyTorch} \cite{anselPyTorch2Faster2024} and \texttt{NumPy} \cite{Harris:2020xlr} for the autoencoders and the numerical back-end.
    \texttt{LISANode} \cite{bayleLISANode2022} was used to generate realistic LISA noise in Appendix~\ref{sec:intervals-and-downsampling}.
\end{acknowledgments}

\appendix

\section{\label{sec:intervals-and-downsampling}Time intervals and sample rate}
Full simulations at the nominal LISA mission duration and sample rate are viable, but this means to have large amounts of the data -- especially far away from the merger -- that do not carry much information about the underlying GWs.
Since many simulations need to be done, and their output needs to be compressed for NLE to be computationally feasible -- the simulations have been run at a lower sample rate $1/\Delta t$ on a restricted time range around the true time of coalescence $\bar{t}_c$ of our target source.
This appendix describes and justifies our choice of values for $\Delta t$, as well as the times before ($t_b$) and after ($t_a$) the merger that will be simulated.

The main effect of increasing the value of $\Delta t$ is to lower the Nyquist frequency $f_N = 1/2\Delta t$.
Obviously, if our data contains GWs with frequencies higher than $f_N$, we would not be sensitive to them.
Conversely, if we knew the maximum frequency $f_\mathrm{max}$ in any GW data we expect to simulate, we know that the GW signal in the range $f_\mathrm{max} < f < f_N$ will be zero.
If we are only interested in the GW signal, we can save some computational time by setting $\Delta t$ so that $f_N$ will be only slightly higher than $f_\mathrm{max}$.
Decreasing $f_N$ in this way also has the convenient side-effect of decreasing the amplitude of the noise in the final time-domain signal -- we are effectively applying a low-pass filter on the data, discarding all the noise with frequency $f > f_N$.

\begin{figure}
    \centering
    \includegraphics[width=\columnwidth]{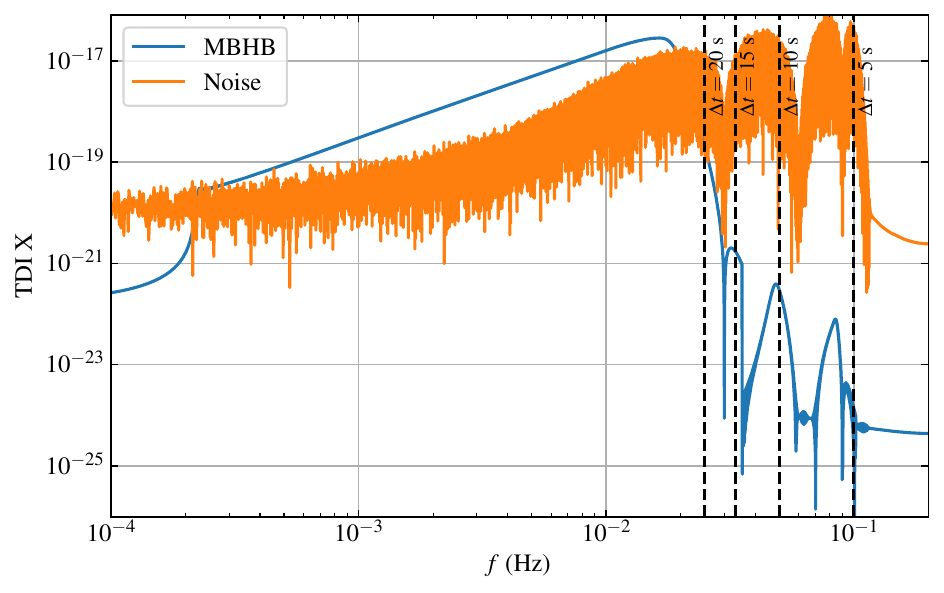}
    \caption{\label{fig:sampling_rate_choice}Frequency-domain view of a simulated low-mass MBHB system (blue) superimposed with non-stationary LISA instrumental noise generated via \texttt{LISANode} (orange).
        The vertical dashed lines represent the Nyquist frequency for different choices of $\Delta t$.}
\end{figure}

In the case of MBHB coalescence events, their merger frequency is inversely proportional to the total mass $M$ of the system.
Indeed, one can perform some back-of-the-envelope estimations using the frequency of the Innermost Stable Circular Orbit (ISCO) of a system \cite{Maggiore:2007ulw},
\begin{equation}
    f^{}_{\mathrm{ISCO}} \simeq \qty{2.2}{\milli\hertz} \left( \frac{\qty{d6}{\solarmass}}{M} \right).
\end{equation}
At the bottom end of our priors ($\mathcal{M}_c = \qty{d5}{\solarmass},\ q=1$), this results in $f_{\mathrm{ISCO}} \simeq \qty{9.58}{\milli\hertz}$; rounding up to $\qty{10}{\milli\hertz}$, the corresponding $\Delta t$ would be $\qty{50}{\second}$.

Of course, the ISCO is the frequency where the inspiral regime ends, but the true merger frequency is higher, and so the required $\Delta t$ will have to be lower.
We therefore determine the merger frequency empirically, by running a frequency-domain simulation of the MBHB system with the lowest $M$ we can reasonably expect to find.
Our value of $\Delta t$ can then be set in such a way that most of the GW signal lies below its Nyquist frequency.
The result can be seen in Fig.~\ref{fig:sampling_rate_choice}, where it is superimposed with \emph{Spritz}-like non-stationary LISA noise generated using the \texttt{LISANode} instrument simulator \cite{bayleLISANode2022}.
From this, we conclude that $\Delta t = \qty{15}{\second}$ is adequate: we can see that frequencies higher than $1/30$ \unit{\hertz} are mostly dominated by instrumental noise, while a lower Nyquist frequency would be dangerously close to the peak frequency of the signal.

For $t_b$ and $t_a$, the inspiral phase is accumulating information over a much longer timescale than the ringdown.
We therefore make $t_a = k_t t_b$, with $k_t < 1$ a hyperparameter denoting the importance of the post-merger timescale compared to the inspiral.
In this work, we have arbitrarily fixed $k_t = 0.25$, meaning that 75\% of the signal will be pre-merger, and 25\% will be post-merger.

\begin{figure}
    \centering
    \includegraphics[width=\columnwidth]{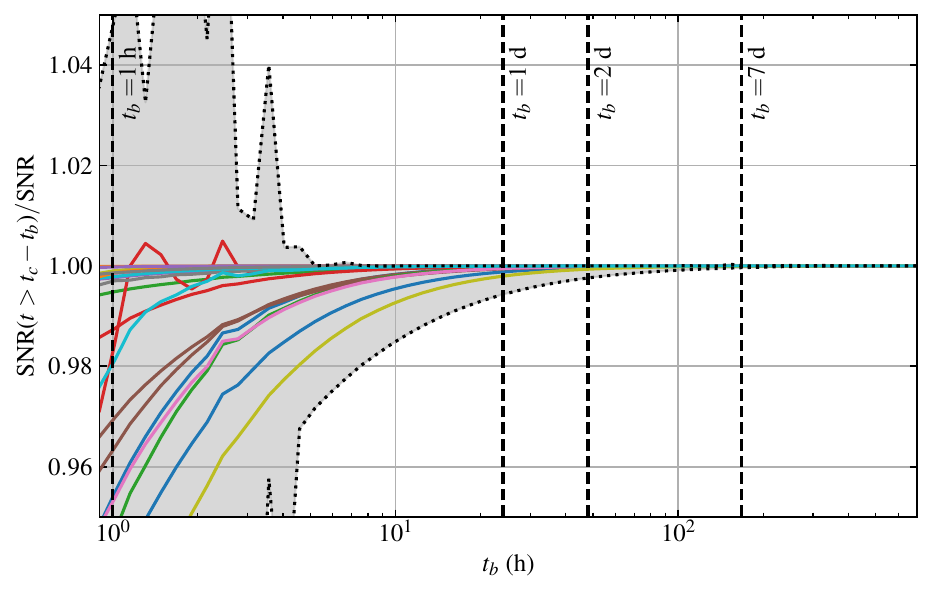}
    \caption{\label{fig:time_trimming}Effect of decreasing observation time on the SNR.
        The vertical axis denotes the SNR (relative to a month-long observation) as the signal is cut to only the last $t_b$ hours (horizontal axis, from right to left).
        The~shaded region denotes the minimum and maximum obtained from a dataset of 20000, with the colored lines serving as illustrative examples from that dataset.
        The dashed vertical lines mark some candidate cutoff times.}
\end{figure}

The only remaining data generation hyperparameter to set is $t_b$. To determine it, the SNR (as defined in Eq.~\eqref{eq:snr}) is used as a proxy for the amount of GW information.
A set of 20000 simulations that have run for a month is generated over broad priors for all parameters.
For each simulation, the total SNR of the waveform is computed.
Then, the start of the time series is gradually cut to emulate a decrease of $t_b$, and the SNR for all waveforms is recomputed at each cut, which is shown in Fig.~\ref{fig:time_trimming}.
As the waveform is cut (reading the figure from right to left), the SNR may begin to decrease, implying a potential loss of information.
From this, we conclude that setting $t_b = \qty{1}{\day}$ (therefore, $T_{\mathrm{obs}} = \qty{30}{\hour}$) seems reasonable: Even in the worst case scenario from our dataset, we find that less than 1\% of the SNR is lost.

\section{\label{sec:noise-psd}LISA noise model}
For the noise PSDs, $S_n^{A, E}(f)$, we use the model that was originally used in the generation of LDC2-b \cite{Babak:2021mhe}.
Here we provide the equations for the model using second-generation TDI and the $A, E$ TDI channels.
The PSD is defined as
\begin{eqnarray}
    S_n^{A,E}(f) & = & 32 \sin^2{x}\sin^2{2x} \left[ (2 + \cos{x}) S^{}_{\mathrm{I}}(f) \right. \nonumber \\[2mm]
    & + &  \left. 2\,(3 + 2 \cos{x} + \cos{2x}) S^{}_{\mathrm{II}}(f) \right]\,,
\end{eqnarray}
where $x = 2 \pi f L/c$, and $L = \qty{2.5d6}{\kilo\meter}$ is the nominal LISA arm length.
The remaining two pieces, $S_{\mathrm{I}}(f)$ and $S_{\mathrm{II}}(f)$ are given by the following expressions:
\begin{equation}
    S^{}_{\mathrm{I}}(f) = S^{}_{\mathrm{oms}} \left( \frac{2 \pi f}{c} \right)^2 \left[ 1 + \left( \frac{\qty{2}{\milli\hertz}}{f} \right)^4 \right]\,,
\end{equation}
\begin{equation}
    S^{}_{\mathrm{II}}(f) = \frac{S_{\mathrm{acc}}}{(2\pi c f)^2} \left[ 1 + \left( \frac{\qty{0.4}{\footnotesize \milli\hertz}}{f} \right)^2 \right] \left[ 1 + \left( \frac{f}{\qty{8}{\footnotesize \milli\hertz}} \right)^4 \right]\,,
\end{equation}
where the quantities $S_{\mathrm{oms}}$ and $S_{\mathrm{acc}}$, correspond to the amplitude of the noise from the Optical Metrology System and from differential acceleration noise respectively.
A complete analysis of LISA data would treat these as unknown parameters and fit these parameters to the residual noise, but for the scope of this work we simply take the values that were used in the production of LDC2-b: $\sqrt{S_{\mathrm{oms}}} = \qty{7.9}{\pico\meter\per\sqrt{\hertz}}$ and $\sqrt{S_{\mathrm{acc}}} = \qty{2.4}{\femto\meter\per\second\squared\per\sqrt{\hertz}}$.
In practice, we use the implementations of these functions that are already present in the LDC software \cite{LDCSoftware}.

\section{\label{sec:ae-architecture}Autoencoder architectures}
Table~\ref{tab:ae-architecture} shows the architecture used in the autoencoders trained in Sec.~\ref{sec:autoencoders} and the SNL-AE analyses of Sec.~\ref{sec:results}.
The dimensions in the third column of Table~\ref{tab:ae-architecture} follow the notation [number of channels, size of each channel].
Layer 5 reshapes all the channels into a single one, so only one dimension is needed.
The 1D convolutional layers are denoted as conv(number of output channels, kernel size, stride).
The max pooling layers are parametrized only by a single number, which is both the kernel size and the stride: a MaxPool(2) layer will thus halve the length of each channel.
In the last two layers, FC stands for fully-connected layers, and the parameter there is the number of neurons in each layer.

\begin{table}
    \centering
    \begin{tabular}{lcc}
        \hline
                                 & $E_{\vecg{\psi}}(\vec{x})$ & Dimensions                   \\
        \midrule
        Input                    & ---                        & [2, 7200]                    \\
        \hline
        \multirow{3}{*}{Layer 1} & BatchNorm                  & \multirow{3}{*}{[128, 3599]} \\
                                 & conv(128, 9, 2)                                           \\
                                 & ReLU                                                      \\
        \hline
        \multirow{4}{*}{Layer 2} & BatchNorm                  & \multirow{4}{*}{[128, 449]}  \\
                                 & conv(128, 9, 2)                                           \\
                                 & ReLU                                                      \\
                                 & MaxPool(4)                                                \\
        \hline
        \multirow{4}{*}{Layer 3} & BatchNorm                  & \multirow{4}{*}{[128, 111]}  \\
                                 & conv(128, 5, 2)                                           \\
                                 & ReLU                                                      \\
                                 & MaxPool(2)                                                \\
        \hline
        \multirow{2}{*}{Layer 4} & BatchNorm                  & \multirow{2}{*}{[32, 111]}   \\
                                 & conv(32, 1, 1)                                            \\
        \hline
        Layer 5                  & Flatten                    & [3552]                       \\
        \hline
        \multirow{2}{*}{Layer 6} & FC(512)                    & \multirow{2}{*}{[512]}       \\
                                 & ReLU                                                      \\
        \hline
        Layer 7                  & FC(128)                    & [128]                        \\
        \hline
    \end{tabular}
    \caption{\label{tab:ae-architecture}Architecture of the encoder $E_{\vecg{\psi}}$ half of the autoencoders trained in this work (see Sec.~\ref{sec:autoencoders}). The dimensionality of the data after each layer is also shown. The architecture for the decoder part is symmetric. See the text of Appendix~\ref{sec:ae-architecture} for details.}
\end{table}

The decoder half $D_{\vecg{\xi}}$ of the autoencoder follows a symmetric architecture to the encoder, so the hyperparameters can be read off Table~\ref{tab:ae-architecture} starting from Layer 6 and going backwards.
The flattening of Layer 5 becomes an unflattening, stacking the flat data into 32 channels.
The convolutional layers become transpose 1D convolutional layers, which functionally serve as the inverse of convolutional layers.
The number of output channels that needs to be used is the number of channels read from the third column of the previous layer.
The max pooling operations become nearest-neighbor upsampling operations: the inverse of a MaxPool(2) will essentially return each data point repeated twice.

It should be noted that convolutions with stride greater than 1 may map different input dimensions to the same output dimension.
In turn, the transposed convolution operations have some ambiguity in the output dimensionality, which can be changed by artificially zero-padding the input on one side.
We have chosen not to do this, which results in decoder outputs of shape $[2\,,\, 7257]$, slightly longer than the input.
To make the decoder output dimension match the encoder input, we symmetrically crop the extra data.


\providecommand{\href}[2]{#2}\begingroup\raggedright\endgroup

\end{document}